\documentclass[aps,prd,twocolumn,nofootinbib,superscriptaddress,altaffilletter,floatfix]{revtex4-1}

\usepackage{graphicx}
\usepackage{float}
\usepackage{amssymb}
\usepackage{amsmath}
\usepackage{longtable}
\usepackage{color}
\usepackage{hyperref}

\usepackage{soul}

\newcommand{\Amp}{\mathcal{A}}
\newcommand{\Dop}{\lambda}
\newcommand{\uCoord}{u}
\newcommand{\DopSpace}{\mathcal{P}}
\newcommand{\sig}{{\mathrm{s}}}
\newcommand{\templ}{\mathrm{t}}
\newcommand{\cosi}{\cos\iota}
\newcommand{\phiO}{\phi_0}
\newcommand{\hO}{h_0}

\newcommand{\coh}[1]{\widetilde{#1}}
\newcommand{\inc}[1]{\widehat{#1}}
\newcommand{\Fstat}{\mathcal{F}}
\newcommand{\Fcoh}{\coh{\Fstat}}
\newcommand{\Finc}{\inc{\Fstat}}
\newcommand{\iSeg}{\ell}

\newcommand{\expect}[1]{E\left[#1\right]}
\newcommand{\avgT}[1]{\left\langle #1 \right\rangle}
\newcommand{\avgN}[1]{\overline{#1}}
\newcommand{\Ord}[1]{\mathcal{O}(#1)}
\newcommand{\modulo}[2]{#1\;\mathrm{mod}\;#2}
\newcommand{\Min}[1]{{#1}_{\mathrm{min}}}
\newcommand{\Max}[1]{{#1}_{\mathrm{max}}}
\newcommand{\MaxMin}[1]{\left|#1\right]}
\newcommand{\var}[1]{\mathrm{var}\left[#1\right]}
\newcommand{\DN}[1]{#1^{\mathrm{DN}}}

\newcommand{\mis}{\mu}
\newcommand{\misCoh}{\coh{\mis}}
\newcommand{\misInc}{\inc{\mis}}

\newcommand{\misF}{\mis_0}
\newcommand{\misFCoh}{\coh{\mis}_0}
\newcommand{\misFInc}{\inc{\mis}_0}

\newcommand{\misMax}{\mis_{\mathrm{max}}}
\newcommand{\misMaxCoh}{\coh{\mis}_{\mathrm{max}}}
\newcommand{\misMaxInc}{\inc{\mis}_{\mathrm{max}}}

\newcommand{\Phase}{\phi}

\newcommand{\tArr}{t_{\mathrm{arr}}}
\newcommand{\tEM}{\tau}
\newcommand{\tRef}{t_{\mathrm{ref}}}
\newcommand{\tPeri}{t_{\mathrm{p}}}
\newcommand{\tSSB}{t_{\mathrm{SSB}}}
\newcommand{\Dt}{\Delta t}
\newcommand{\tAsc}{t_\mathrm{asc}}
\newcommand{\atmid}{{\mathrm{mid}}}
\newcommand{\tMid}{t_\atmid}
\newcommand{\tMidl}{t_{\atmid,\ell}}
\newcommand{\Dma}{\Delta_\mathrm{ma}}
\newcommand{\Dmal}{\Delta_\mathrm{ma,\ell}}
\newcommand{\refine}{\gamma}

\newcommand{\src}{\mathrm{src}}
\newcommand{\vecn}{\vec{n}}
\newcommand{\vecr}{\vec{r}}
\newcommand{\trueanom}{\upsilon}

\newcommand{\freq}{f}
\newcommand{\fdot}{\dot{\freq}}
\newcommand{\fbar}{\bar{\freq}}
\newcommand{\ecc}{e}

\newcommand{\asini}{a_{\mathrm{p}}}			
\newcommand{\argp}{\omega}
\newcommand{\Om}{\Omega}

\newcommand{\Tseg}{\Delta T}
\newcommand{\Tobs}{T_\mathrm{obs}}
\newcommand{\SFT}{{\mathrm{sft}}}
\newcommand{\Tsft}{T_\SFT}
\newcommand{\Nseg}{N}
\newcommand{\Ndet}{N_{\mathrm{det}}}
\newcommand{\Resamp}{{\mathrm{FFT}}}

\newcommand{\psim}{\psi_{\mathrm{m}}}

\newcommand{\Hz}{\mathrm{Hz}}
\newcommand{\Sec}{\mathrm{s}}
\newcommand{\Hours}{\mathrm{h}}
\newcommand{\Days}{\mathrm{d}}

\newcommand{\Nt}{\mathcal{N}}
\newcommand{\NtInc}{\inc{\Nt}}
\newcommand{\NtCoh}{\coh{\Nt}}
\newcommand{\NtPerDim}[1]{\Nt_{#1}}

\newcommand{\LongSeg}{\mathrm{LS}}
\newcommand{\ShortSeg}{\mathrm{SS}}

\newcommand{\Cost}{\mathcal{C}}
\newcommand{\Ctot}{\Cost_{\mathrm{tot}}}
\newcommand{\CInc}{\inc{\Cost}}
\newcommand{\CCoh}{\coh{\Cost}}

\newcommand{\pFA}{p_{\mathrm{fa}}}

\newcommand{\relerr}{\varepsilon}

\newcommand{\SensDepth}[2]{\mathcal{D}^{#1}_{#2}}
\newcommand{\Sn}{S_{\mathrm{n}}}
\newcommand{\ScoX}{\text{ScoX1}}

\newcommand{\cR}{\mathcal{R}}
\newcommand{\Zn}[1]{\mathbb{Z}_{#1}}
\newcommand{\Ans}[1]{A^{*}_{#1}}
\newcommand{\thick}[1]{\theta_{#1}}

\begin{document}

\title{
Directed searches for continuous gravitational waves from binary systems:
parameter-space metrics and optimal Scorpius~X-1 sensitivity}


\author{Paola~Leaci}
\altaffiliation{paola.leaci@roma1.infn.it}
\affiliation{Max-Planck-Institut f\"ur Gravitationsphysik, Albert-Einstein-Institut, D-14476 Golm, Germany}
\affiliation{Dip. di Fisica, Universit\`{a} di Roma ``Sapienza'', P.le A. Moro, 2, I-00185 Rome, Italy}
\author{Reinhard~Prix}
\altaffiliation{reinhard.prix@aei.mpg.de}
\affiliation{Max-Planck-Institut f\"ur Gravitationsphysik, Albert-Einstein-Institut, D-30167 Hannover, Germany}
\date{\today}

\begin{abstract}
  We derive simple analytic expressions for the (coherent and semi-coherent) phase metrics of
  continuous-wave sources in low-eccentricity binary systems, for the two regimes of long and short
  segments compared to the orbital period.
  The resulting expressions correct and extend previous results found in the literature.
  We present results of extensive Monte-Carlo studies comparing metric mismatch predictions against
  the measured loss of detection statistic for binary parameter offsets.
  The agreement is generally found to be within $\sim10\%-30\%$.
  As an application of the metric template expressions, we estimate the optimal achievable
  sensitivity of an Einstein@Home directed search for Scorpius~X-1, under the assumption of
  sufficiently small spin wandering.
  We find that such a search, using data from the upcoming advanced detectors, would be able to beat
  the torque-balance level~\cite{wagoner84:_gravit,bildsten98:_gravit} up to a frequency
  of $\sim500 - 600\,\Hz$, if orbital eccentricity is well-constrained, and up to a frequency of
  $\sim160-200\,\Hz$ for more conservative assumptions about the uncertainty on orbital eccentricity.
\end{abstract}

\maketitle

\section{Introduction}
\label{Intro}

Continuous gravitational waves (CWs) are a promising class of signals for the second-generation
detectors currently under construction: advanced LIGO (aLIGO) \cite{AdvLIGORef:2010},
advanced Virgo \cite{AdvVirgoRef:2009} and KAGRA\cite{KAGRA_Ref:2012}.
These signals would be emitted by spinning neutron stars (NSs)
subject to non-axisymmetric deformations, such as quadrupolar deformations (``mountains''),
unstable oscillation modes (e.g.\ r-modes) or free precession. For a general review of CW sources
and search methods, see for example \cite{prix06:_cw_review}.

A particularly interesting type of potential CW sources are NSs in low-mass X-ray binaries (LMXBs),
with Scorpius~X-1 being its most prominent representative \cite{Watts:2008qw}.
The accretion in these systems would be expected to have spun up the NSs to the maximal rotation
rate of $\gtrsim 1000\,\Hz$~\cite{Cook:1994a}. All the observations performed to date, however, show they only spin at several hundred
$\Hz$, so there seems to be something limiting the accretion-induced spin-up.
A limiting mechanism that has been suggested is the emission of gravitational
waves~\cite{wagoner84:_gravit,bildsten98:_gravit,NonAxNS2}, which would result in steady-state
CW emission where the accretion torque is balanced by the radiated angular momentum.
For a discussion of alternative explanations, see ~\cite{bildsten98:_gravit, NonAxNS2}.
The resulting torque-balance CW amplitude increases with the observed X-ray flux (independent of the
distance of the system, e.g.\ see Eq.(4) in \cite{bildsten98:_gravit}), therefore the X-ray
brightest LMXB Scorpius~X-1 is often considered the most promising CW source within this category
\cite{Watts:2008qw,ScoX1:MDC1}.

Several searches for CW signals from Scorpius~X-1 have been performed (without any detections) on data from
initial LIGO~\cite{Abbott:2006vg,Aasi:2014zwg,2014arXiv1412.0605T}, and several new pipelines have been developed and are
currently being tested in a Scorpius~X-1 Mock Data Challenge (MDC)~\cite{ScoX1:MDC1}.
So far, all current methods fall into either one of two extreme cases: highly coherent with a
short total time baseline (6\,hours in~\cite{Abbott:2006vg}, 10\,days
in~\cite{2014arXiv1412.0605T}), or highly incoherent with a long
total baseline but very short ($\sim$hours) coherent segments~\cite{Aasi:2014zwg,Chung:2011da,Aasi:2014sda}.

In order to increase sensitivity beyond these methods, and to be able to effectively absorb
large amounts of computing power (such as those provided by Einstein@Home
\cite{Einstweb,Aasi:2012fw}), it is necessary to extend the search approach into the realm of
general long-segment semi-coherent methods, by stacking coherent $\Fstat$-statistic segments
\cite{Brady:1998nj,HierarchP2,Prix:2012yu}.
Such methods have already been employed over the past years for both directed and all-sky searches for
CWs from isolated NSs \cite{Pletsch:2009uu,Aasi:2012fw,2013PhRvD..88j2002A}, but they have not yet
been extended to the search for CWs from binary systems.

Key ingredients required for building a semi-coherent ``StackSlide'' search are the coherent and
semi-coherent parameter-space metrics \cite{Owen:1995tm,Balasubramanian:1995bm,Brady:1997ji,Brady:1998nj}.
The coherent binary CW metric was first analyzed in \cite{Dhurandhar:2000sd}, and this was further
developed and extended to the semi-coherent case in \cite{Messenger:2011rg}.
Here we will largely follow the approach of \cite{Messenger:2011rg}, considering only
low-eccentricity orbits, and focusing on two different regimes of either long coherent segments, or
short segments compared to the orbital period, both of which admit simple analytic results.

\subsubsection*{Plan of this paper}
\label{sec:plan-this-paper}

Sec.~\ref{Sec:Background} provides a general introduction of the concepts and notation
of semi-coherent StackSlide methods, parameter-space metric and template banks.
Secs.~\ref{sec:binary-cw-phase}--\ref{sec:number-templates} build on the work of
\cite{Messenger:2011rg}, rectifying some of the results and extending them to the case of a general
orbital reference time.
In Sec.~\ref{softwaresim} we present the results of extensive Monte-Carlo software-injection tests comparing
the metric predictions against the measured loss in statistics due to parameter-space offsets.
Sec.~\ref{sec:sco-x1-sensitivity} applies the theoretical template-bank counts to compute
optimal StackSlide setups for a directed search for Scorpius~X-1 and estimates the resulting achievable
sensitivity. Finally, Sec.~\ref{sec:conc} presents concluding remarks.

\textit{Notation.} Throughout the paper we denote a quantity $Q$ as $\coh{Q}$ when referring to the
coherent case, and as $\inc{Q}$ when referring to the semi-coherent case.

\section{Background}
\label{Sec:Background}

The strain $h$ in a given detector due to a continuous gravitational wave is a scalar function
$h(t; \Amp,\Dop)$, where $t$ is the time at the detector.
The set of signal parameters $\Amp$ denotes the four \emph{amplitude parameters}, namely the overall
amplitude $\hO$, polarization angles $\cosi$ and $\psi$, and the initial phase $\phiO$. The set of
\emph{phase-evolution parameters} $\Dop$ consist of all the remaining signal parameters affecting
the time-evolution of the CW phase at the detector, notably the signal frequency $\freq$, its frequency
derivatives (also known as ``spindown'' terms), sky-position of the source, and binary orbital parameters.
We will look in more detail at the phase model in Sec.~\ref{sec:binary-cw-phase}.

The observed strain $x(t)$ in the detector affected by additive Gaussian noise $n(t)$ can
be written as $x(t) = n(t) + h(t;\Amp,\Dop)$, and the \emph{detection problem} consists in
distinguishing the (pure) ``noise hypothesis'' of $\hO = 0$ from the ``signal hypothesis'' $\hO > 0$.

\subsection{Coherent detection statistic}
\label{sec:coher-detect-stat}
As shown in \cite{Jaranowski:1998qm}, matched-filtering the data $x(t)$ against a template
$h(t;\Amp,\Dop)$, and analytically maximizing over the unknown amplitude parameters $\Amp$, results in
the coherent statistic $\Fcoh(x;\Dop)$, which only depends on the data $x$ and on the
template phase-evolution parameters $\Dop$.
This statistic follows a (non-central) $\chi^2$ distribution with $4$ degrees of freedom, and a
non-centrality parameter $\coh{\rho}^2(\Amp,\Dop_\sig;\Dop)$, which depends on the signal parameters
$\{\Amp,\Dop_\sig\}$ and the template parameters $\Dop$. The quantity $\coh{\rho}$ (which depends
linearly on $\hO$) is generally referred to as the signal-to-noise ratio (SNR) of the coherent
$\Fcoh$-statistic.
The expectation value of this statistic over noise realizations is given by~\cite{Jaranowski:1998qm}
\begin{equation}
  \label{eq:3}
  \expect{2\Fcoh(x;\Dop)} = 4 + \coh{\rho}^2(\Amp,\Dop_\sig;\Dop)\,.
\end{equation}
In the case of unknown signal parameters $\Dop_\sig\in\DopSpace$ within some parameter space
$\DopSpace$, the number of required template $\Dop_\templ$ that need to be searched is typically
impractically large.
The maximal achievable sensitivity by this coherent statistic $\Fcoh(x;\Dop)$ is therefore severely
limited \cite{Brady:1997ji} by the required computing cost.
It was shown that \emph{semi-coherent} statistics generally result in better
sensitivity at equal computing cost (e.g.\ see \cite{Brady:1998nj,Prix:2012yu}).

\subsection{Semi-coherent detection statistic}
\label{sec:semi-coher-detect}
Here we focus on one particular semi-coherent approach, sometimes referred to as
StackSlide, which consists of dividing the total amount of data $\Tobs$ into $\Nseg$ segments of
duration $\Tseg$, such that $\Tobs = \Nseg\,\Tseg$ (in the ideal case of gapless data).
The coherent statistic $\Fcoh_\iSeg(x;\Dop)$ is then computed over all segments
$\iSeg = 1\ldots\Nseg$, and combined incoherently by summing\footnote{This form of the
  ``ideal'' StackSlide statistic is not directly used for actual searches. For reasons of
  computational cost, the coherent statistics across segments are computed on coarser template
  banks, and are combined on a fine template bank by summing across segments using interpolation on
  the coarse grids. This is discussed in detail in \cite{Prix:2012yu}, but is not relevant for the
  present investigation.}
\begin{equation}
  \label{eq:2}
  \Finc(x;\Dop) \equiv \sum_{\iSeg=1}^\Nseg \Fcoh_\iSeg(x;\Dop)\,.
\end{equation}
This statistic follows a (non-central) $\chi^2$-distribution with $4\Nseg$ degrees of
freedom and a non-centrality parameter $\inc{\rho}^2(\Amp,\Dop_\sig;\Dop)$, which is found to be
given by
\begin{equation}
  \label{eq:1}
  \inc{\rho}^2 = \sum_{\iSeg=1}^\Nseg \coh{\rho}_\iSeg^2\,,
\end{equation}
in terms of the per-segment coherent SNRs $\coh{\rho}_\iSeg$. Note that (contrary to the
coherent case) $\inc{\rho}$ can \emph{not} be regarded as a semi-coherent SNR.
The expectation for $\inc{\Fstat}$ is
\begin{equation}
  \label{eq:4}
  \expect{2\Finc(x;\Dop)} = 4\Nseg + \inc{\rho}^2(\Amp,\Dop_\sig;\Dop)\,.
\end{equation}

\subsection{Template banks and metric mismatch}
\label{sec:templ-banks-metr}
In order to systematically search a parameter space $\DopSpace$ using a statistic $\Fstat(x;\Dop)$
(which here can refer either to the coherent $\Fcoh$ or the semi-coherent $\Finc$), we need to select
a finite sampling $\{\Dop_\templ\}\subset\DopSpace$ of the parameter space, commonly referred to as
a template bank, and compute the statistic over this set of templates, i.e.\ $\{\Fstat(x;\Dop_\templ)\}$.

A signal with parameters $\Dop_\sig\in\DopSpace$ will generally not fall on an exact template, and
we therefore need to characterize the loss of detection statistic $\Fstat$ as a function of the
offset $\delta\Dop = \Dop - \Dop_\sig$ from a signal.
This is generally quantified using the \emph{expected} statistic $\expect{2\Fstat}$ of
Eqs.~\eqref{eq:3} and \eqref{eq:4}, namely (by removing the bias $4\Nseg$) as the relative loss in
the non-centrality $\rho^2$:
\begin{equation}
  \label{eq:5}
  \misF \equiv \frac{ \rho^2(\Amp,\Dop_\sig;\Dop_\sig) - \rho^2(\Amp,\Dop_\sig;\Dop)}{\rho^2(\Amp,\Dop_\sig;\Dop_\sig)}\,,
\end{equation}
which defines the measured $\Fstat$-statistic \emph{mismatch} function. This mismatch has a global
minimum of $\misF=0$ at $\Dop=\Dop_\sig$ and is bounded within $\misF\in[0,1]$.

We now use two standard approximations: \textit{(i)} Taylor-expand this up to second
order in small offsets $\delta\Dop$, and \textit{(ii)} neglect the dependence on the amplitude parameters
$\Amp$ (the effect of which was analyzed in detail in \citep{Prix:2006wm}), which leads to the
well-known \emph{phase metric} $g_{ij}$, namely
\begin{align}
  \misF(\Amp,\Dop_\sig;\Dop) &\approx \mis(\Dop_\sig;\Dop)\,,\quad\text{where} \label{eq:6a}\\
  \mis(\Dop_\sig;\Dop) &\equiv g_{ij}(\Dop_\sig)\,\delta\Dop^i\delta\Dop^j\,, \label{eq:6b}
\end{align}
with $g_{ij}$ a positive-definite symmetric matrix, and implicit summation over repeated
indices $i,j$ is used.
The quality of this approximation will be quantified in a Monte-Carlo study in Sec.~\ref{softwaresim}.
The \emph{metric mismatch} $\mis$ has a global minimum of $\mis=0$ at $\Dop=\Dop_\sig$ and (contrary
to $\misF$) is semi-unbounded, i.e.\ $\mis\in[0,\infty)$. Previous studies testing $\mis$ against
$\misF$ \cite{Prix:2006wm,Wette:2013wza,Wette:2014a} for all-sky searches have shown that the
phase-metric approximation Eq.~\eqref{eq:6a} works reasonably well for observation times
$\Tobs\gtrsim\Ord{1\,\mathrm{day}}$ and for mismatch values up to $\mis\lesssim 0.3 - 0.5$. Above this
mismatch regime, the metric mismatch $\mis$ increasingly \emph{over-estimates} the actual loss
$\misF$.
For example, in \cite{Prix:2006wm} an empirical fit for this behavior was given as
$\misF(\mis) \approx \mis - 0.38\,\mis^2$.

The metric formalism was first introduced in \cite{Owen:1995tm,Balasubramanian:1995bm} (in the context
of compact binary coalescence signals), and was later applied to the CW problem in \cite{Brady:1997ji}, where it
was shown that the \emph{coherent} phase metric $\coh{g}_{ij}$ can be expressed explicitly as
\begin{equation}
  \label{eq:7}
  \coh{g}_{ij}(\Dop) = \avgT{\partial_i\Phase(\Dop) \, \partial_j\Phase(\Dop)} - \avgT{\partial_i\Phase(\Dop)}\avgT{\partial_j\Phase(\Dop)}\,,
\end{equation}
where
$\partial_i\Phase \equiv {\partial \Phase}/{\partial \Dop^i}$, with $\Phase(t;\Dop)$ the CW
signal phase, and $\avgT{\ldots}$ denotes the time average
over the coherence time $T$, i.e.\ $\avgT{Q} \equiv \frac{1}{T}\int_{t_0}^{t_0+T}Q(t)\,dt$.

The corresponding \emph{semi-coherent} phase metric $\inc{g}_{ij}$ was first studied in
\cite{Brady:1998nj} and was found to be expressible as the average over all per-segment coherent
phase metrics\footnote{This assumes constant signal SNR
$\coh{\rho}_\iSeg(\Amp,\Dop_\sig;\Dop_\sig)$ over all segments $\iSeg$, which should be a good
approximation for stationary noise and sufficiently long segments, such that diurnal antenna-pattern
variations have averaged out.})
$\coh{g}_{\iSeg,ij}$, i.e.
\begin{equation}
  \label{eq:8}
  \inc{g}_{ij}(\Dop) = \avgN{\coh{g}_{ij}(\Dop)}\,,
\end{equation}
where we defined the average operator over segments as
\begin{equation}
  \label{eq:67}
  \avgN{Q} \equiv \frac{1}{\Nseg}\sum_{\iSeg=1}^{\Nseg} Q_\iSeg\,.
\end{equation}
The metric is useful for constructing template banks, by providing a simple criterion
for how
``close'' we need to place templates in order to limit the maximal (relative) loss in detection
statistic, which can be written as
\begin{equation}
  \label{eq:10}
  \max_{\Dop_\sig}\, \min_{\Dop_\templ} \mis(\Dop_\sig;\Dop_\templ) \le \misMax\,.
\end{equation}
This states that the worst-case mismatch to the closest template over the whole template bank should be
bounded by a maximal value $\misMax$.
Note that each template $\Dop_\templ$ ``covers'' a parameter-space volume given by
\begin{equation}
  \label{eq:11}
  \mis(\Dop_\templ;\Dop) = g_{ij}(\Dop_\templ)\,\delta\Dop^i\,\delta\Dop^j \le \misMax\,,
\end{equation}
which describes an $n$-dimensional ellipsoid, where $n$ is the number of
parameter-space dimensions. The template bank can therefore be thought of as a covering of the whole
parameter space with such template ellipsoids, such that no region of $\DopSpace$ remains
uncovered.
One can show \cite{Brady:1997ji,Prix:2007ks,Messenger:2008ta} that the resulting number of templates
$\Nt$ in such a template bank is expressible as
\begin{equation}
  \label{eq:NumTemplates}
  \Nt = \thick{n} \, \misMax^{-n/2}\, \int_{\DopSpace} \,\sqrt{\det g(\Dop)}\,d^{n}\Dop ,
\end{equation}
where $n$ is the number of template-bank dimensions, and $\thick{n}$ is the \emph{center density}
(also known as \emph{normalized thickness}) of the covering lattice, which quantifies the number of
lattice points per (metric) volume for a unit mismatch.
The center density is a geometric property of the lattice, and for the typical covering lattices
used here is given by~\cite{Prix:2007ks}
\begin{equation}
  \label{eq:77}
  \thick{n} =
  \begin{cases}
    2^{-n}\,n^{n/2} & \text{for }\Zn{n}\,,\\
    \sqrt{n+1}\,\left[\frac{n(n+2)}{12(n+1)}\right]^{n/2} & \text{for } \Ans{n}\,.
  \end{cases}
\end{equation}
An important caveat for using Eq.~\eqref{eq:NumTemplates} is that the metric determinant and integration must
only include \emph{fully resolved} template-bank dimensions, which require more than one
template to cover the extent $\Delta\Dop^i$ of the parameter space along that dimension.
We can estimate the per-dimension template extents at given maximal mismatch $\misMax$
as\footnote{Note that the extra factor of $2$ here compared to \cite{Watts:2008qw} is required to
  account for the total extent of the ellipse, not just the extent from the center.
  This can also be seen by considering that the center density $\thick{\Zn{1}} = 1/2$ in
  Eq.~\eqref{eq:NumTemplates} for the 1-dimensional case.
}
\begin{equation}
  \label{eq:82}
  \delta\Dop^i = 2\sqrt{\misMax\,[g^{-1}]^{ii}}\,,
\end{equation}
where $[g^{-1}]^{ij}$ are the elements of the inverse matrix to the metric $g_{ij}$.
We can therefore define the metric template-bank ``thickness'' along dimension $i$ in terms of the
corresponding effective number of templates along that direction as
\begin{equation}
  \label{eq:89}
  \NtPerDim{\Dop^i} \equiv \frac{\Delta\Dop^i}{\delta\Dop^i}\,.
\end{equation}
For dimensions $\Dop^i$ with $\NtPerDim{\Dop^i}\le 1$ we must exclude this coordinate from the bulk
template-counting formula Eq.~\eqref{eq:NumTemplates}, as it would effectively contribute
``fractional templates'' and thereby incorrectly \emph{reduce} the total template count.


\section{The binary CW phase}
\label{sec:binary-cw-phase}

\subsection{The general phase model}
\label{sec:gener-newt-binary}

The general CW phase model assumes a slowly spinning-down NS with rotation rate $\nu(\tEM)$ and a
quadrupolar deformation resulting in the emission of CWs. The phase evolution can therefore
be expressed as a Taylor series in the NS source frame as
\begin{equation}
  \label{eq:12}
  \Phase^\src(\tEM) = 2\pi\,[ \freq\,(\tEM-\tRef) +
    \frac{1}{2}\fdot(\tEM-\tRef)^2 + \ldots ]\,,
\end{equation}
where $\tRef$ denotes the reference time and $\freq,\fdot,\ddot{\freq},\ldots$ are the CW frequency
and spindown parameters. These are defined as
\begin{equation}
  \label{eq:31}
  \freq^{(k)} \equiv \sigma\,\left.\frac{d^k \nu(\tEM)}{d \tEM^k}\right|_{\tEM=\tRef}\,,
\end{equation}
where $\sigma$ is a model-dependent constant relating the instantaneous CW frequency
$\freq(\tEM)$ to the NS spin frequency $\nu(\tEM)$. For example,
for a quadrupolar deformation rotating rigidly with the NS (``mountain'') we have
$\sigma=2$~\cite{bildsten98:_gravit,NonAxNS2,Cutler:2002nw,Melatos:2005ez,Owen:2005fn},
while for r-modes $\sigma\approx4/3$~\cite{bildsten98:_gravit,Owen:1998xg,Andersson:1998qs} and for
precession $\sigma\approx 1$~\cite{Jones:2001yg,VanDenBroeck:2004wj}.

In order to relate the CW phase in the source frame to the phase $\Phase(\tArr)$ in the detector frame
needed for Eq.~\eqref{eq:7}, we need to relate the wavefront detector arrival time $\tArr$ to its
source emission time $\tEM$, i.e.\ $\tEM(\tArr)$, such that $\Phase(\tArr) = \Phase^\src(\tEM(\tArr))$.
Neglecting relativistic wave-propagation effects\footnote{These effects are taken into account in
  the actual matched-filtering search codes, but are negligible for the calculation of the metric.}
(such as Einstein and Shapiro delays, e.g.\ see \cite{2006MNRAS.369..655H,2006MNRAS.372.1549E} for
more details), we can write this as
\begin{equation}
  \label{eq:13}
  \tEM(\tArr) = \tArr + \frac{\vecr(\tArr)\cdot\vecn}{c} - \frac{d}{c} - \frac{R(\tEM)}{c}\,,
\end{equation}
where $\vecr$ is the vector from solar-system barycenter (SSB) to the detector, $c$
is the speed of light, $d$ is the (generally unknown) distance between the SSB and
the binary barycenter (BB), $R$ is the radial distance of the CW-emitting
NS from the BB along the line of sight, where $R>0$ means the NS is
further away from us than the BB, and $\vecn$ is the unit vector pointing from the
SSB to the source. In standard equatorial coordinates with right ascension $\alpha$
and declination $\delta$, the components of the unit vector $\vecn$ are given by
$\vecn=(\mathrm{cos}\,\alpha \, \mathrm{cos}\,\delta,\ \mathrm{sin}\,\alpha \, \mathrm{cos}\,\delta,\ \mathrm{sin} \,\delta)$.

The projected radial distance $R$ along the line of sight can be expressed
(e.g.\ see \cite{2005ormo.book.....R}) as
\begin{equation}
  \label{eq:17}
  R =y\,\sin{}i\,\sin(\argp+\trueanom)\,,
\end{equation}
where $y$ is the distance of the NS from the BB,
$i$ is the inclination angle between the orbital plane and the sky, $\argp$ is the \emph{argument of
periapse}, and $\trueanom$ is the \emph{true anomaly} (i.e.\ the angle from the periapse to the current NS
location around the BB). We further approximate the orbital motion as a pure Keplerian ellipse,
which can be described as
\begin{equation}
  \label{eq:14}
  y(\trueanom) = \frac{a\,(1 - \ecc^2)}{1 + \ecc\,\cos\trueanom}\,,
\end{equation}
in terms of the semi-major axis $a$ and the eccentricity $\ecc$.
The ellipse can be written equivalently in terms of the \emph{eccentric anomaly} $E$, namely
\begin{equation}
  \label{eq:15}
  y(E) = a ( 1 - \ecc\,\cos E)\,,
\end{equation}
and the dynamics is described by Kepler's equation, namely
\begin{equation}
  \label{eq:16}
  \tEM - \tPeri = \frac{P}{2\pi}\left( E - \ecc\,\sin E\right)\,,
\end{equation}
which provides a (transcendental) relation for $E(\tau)$.
Combining Eqs.~\eqref{eq:17},~\eqref{eq:14} and~\eqref{eq:15}, we can rewrite the projected radial
distance $R$ in terms of $E$, namely
\begin{equation}
  \label{eq:18}
  \frac{R}{c} = \asini\left[ \sin\argp ( \cos E - \ecc) + \cos\argp\sin E \sqrt{1 - \ecc^2} \right]\,,
\end{equation}
where we defined $\asini\equiv a\sin{}i/c$.
Combining this with Eq.~\eqref{eq:16} fully determines (albeit only implicitly) the
functional relation $R(\tEM)$ required for the timing model of Eq.~\eqref{eq:13}.

Dropping the unknown distance $d$ to the BB (which is equivalent to re-defining the intrinsic
spindown parameters), and defining the SSB wavefront arrival time $\tSSB$ as
\begin{equation}
  \label{eq:19}
  \tSSB(\tArr;\vecn) \equiv \tArr + \frac{\vecr(\tArr)\cdot\vecn}{c}\,,
\end{equation}
we can rewrite the timing relation Eq.~\eqref{eq:13} as
\begin{equation}
  \label{eq:20}
  \tEM(\tSSB) = \tSSB - \frac{R(\tEM)}{c}\,.
\end{equation}

Here we are interested only in binary systems with \emph{known} sky-position
$\vecn$, and we can therefore change integration variables from $t$ to $\tSSB$ in
the phase for the metric integration of Eq.~\eqref{eq:7}.
Furthermore, given that the Earth's R\o{}mer delay is bounded by $|\vecr\cdot\vecn/c|\lesssim500\,$s,
and $|d\tSSB/dt|\lesssim 10^{-4}$, this difference will be negligible for the metric and can be
dropped. This is equivalent to effectively placing us into the SSB, which is always
possible for known $\vecn$. In order to simplify the notation, we now simply write $t \equiv \tSSB$.

Plugging this timing model into the phase of Eq.~\eqref{eq:12}, we obtain\footnote{The
  sign on $R$ (and equivalently on $\asini$) here agrees with Eq.(2.26) in \cite{Blandford:1976},
  but differs from Eq.(2) in \cite{Messenger:2011rg}, which is incorrect.}
\begin{equation}
  \label{eq:21}
  \Phase(t) \approx 2\pi\left[ \freq \left(\Dt - \frac{R}{c}\right) + \frac{1}{2}\fdot \left(\Dt - \frac{R}{c}\right)^2 + \ldots\right]\,,
\end{equation}
with $\Dt \equiv t - \tRef$.
The binary systems we are interested in have semi-major axis $\asini$ of order
$\Ord{1-10}\,$s, and binary periods $P$ of order of several hours. Hence, the change in $E$,
and therefore $R(E)$, during the time $R/c$ will be negligible, and so we can approximate
$E(\tEM) \approx E(t)$, namely
\begin{equation}
  \label{eq:22}
  t - \tPeri \approx \frac{P}{2\pi}\left( E - \ecc\,\sin E\right)\,.
\end{equation}
For the purpose of calculating the metric using Eq.~\eqref{eq:7}, we can further approximate the
phase model in the following standard way (e.g.\ see \cite{Jaranowski:1998qm,Wette:2013wza}) as
\begin{equation}
  \label{eq:23}
  \Phase(t) \approx 2\pi\left[
    (\freq\Dt + \frac{1}{2!}\fdot\Dt^2 + \ldots) - \frac{R(t)}{c} \fbar
  \right]\,,
\end{equation}
where we expanded the factors $(\Dt-R/c)^k$ and kept only the leading-order terms in
$\Dt$, keeping in mind that $\Dt\sim T \gtrsim\Ord{\mathrm{days}}$ and
$R/c\sim\Ord{\mathrm{seconds}}$, where $T$ denotes the coherence time.
In the last term we replaced the instantaneous intrinsic CW frequency as a function of $t$ by a
constant parameter $\fbar$, namely we approximated $\fbar \approx (\freq+\fdot\Dt+\ldots)$. This
``frequency scale'' $\fbar$ of the signal could be chosen as the average (or the largest) intrinsic CW
frequency of this phase model over the coherence time $T$. Given that this only enters as a scale
parameter in the metric, and the changes of intrinsic frequency over the observation time will
generally be small, this introduces only a negligible difference.

\subsection{The small-eccentricity approximation ($\ecc\ll1$)}
\label{sec:small-eccentr-appr}

We follow the approach of \cite{Messenger:2011rg} and consider only the
\emph{small-eccentricity limit} of the metric. Hence, by Taylor expanding
Eqs.~\eqref{eq:18} and \eqref{eq:22} up to leading order in $\ecc$, i.e.\ inserting
$E(t) = E_0(t) + \ecc E_1(t) + \ldots$ into Kepler's equation Eq.~\eqref{eq:22}, we obtain
\begin{align}
  \label{eq:24}
  E_0(t) &= \Om ( t - \tPeri )\,,\\
  E_1(t) &= \sin E_0(t)\,,
\end{align}
where $\Om \equiv \frac{2\pi}{P}$ is the \emph{mean} orbital angular velocity.
Plugging this into Eq.~\eqref{eq:18}, we obtain the R\o{}mer delay of the binary to leading order in
$\ecc$ as
\begin{equation}
  \label{eq:25}
  \frac{R}{c} = \asini \left[ \sin\psi(t) + \frac{\kappa}{2}\sin2\psi(t) - \frac{\eta}{2}\cos2\psi(t) \right]\,,
\end{equation}
where a constant term $-3\asini\,\eta/2$ was omitted, which is irrelevant
for the metric.
We use the standard \emph{Laplace-Lagrange} parameters defined as
\begin{align}
   \kappa &\equiv \ecc\,\cos(\argp)\,, \label{eq:26}\\
  \eta  &\equiv \ecc\,\sin(\argp)\, \label{eq:26b} ,
\end{align}
and the mean orbital phase
\begin{equation}
  \label{eq:27}
  \psi(t) \equiv \Om\,(t - \tAsc)\,,
\end{equation}
measured from the \emph{time of ascending node} $\tAsc$, which (for small $\ecc$) is related to $\tPeri$ by \cite{2006MNRAS.369..655H}
\begin{equation}
  \label{eq:28}
  \tAsc \equiv \tPeri - \frac{\argp}{\Om}\,,
\end{equation}
and which (contrary to the time of periapse $\tPeri$), remains well-defined even in the limit of
circular orbits.

The small-eccentricity phase model is therefore parametrized by the 5 binary parameters
$\{\asini,\,\tAsc,\,\Om,\,\kappa,\,\eta\}$ (referred to as the ``ELL1''  model in
\cite{2006MNRAS.369..655H,2006MNRAS.372.1549E}), and in the circular case ($\ecc=0$) this reduces to
the 3 binary parameters $\{\asini,\,\tAsc,\,\Om\}$ with $\kappa=\eta=0$.

\section{The binary CW metric}
\label{sec:binary-cw-phase-1}

\subsection{Phase derivatives}
\label{sec:phase-derivatives}

Following \cite{Dhurandhar:2000sd,Messenger:2011rg} we restrict our
investigation to (approximately) constant-frequency CW signals, i.e.\ $\freq^{(k>0)}=0$.
This is motivated by the assumed steady-state torque-balance situation in LMXBs,
which are our main target of interest. However, the corresponding fluctuations in the accretion rate
are expected to cause some stochastic frequency drift, and one will therefore need to be careful
to restrict the maximal coherence time in order to limit the frequency resolution. This will be
discussed in more detail in the application to Scorpius~X-1 in Sec.~\ref{sec:sco-x1-sensitivity}.
The total phase-evolution parameter space considered here is therefore spanned by the following 6 coordinates
\begin{equation}
  \label{eq:29}
  \Dop = \{\freq,\,\asini,\,\tAsc,\,\Om,\,\kappa,\,\eta \}\,.
\end{equation}
The small-eccentricity phase model used here can now be explicitly written as
\begin{equation}
  \label{eq:30}
  \frac{\Phase(t;\Dop)}{2\pi} \approx \freq\,\Dt - \fbar \asini \left[ \sin\psi + \frac{\kappa}{2}\sin2\psi - \frac{\eta}{2}\cos2\psi \right]\,,
\end{equation}
with $\psi \equiv \psi(t)$ given by Eq.~\eqref{eq:27}.
The frequency parameter $\fbar$ is treated as a constant in the phase derivative
$\partial_\freq\Phase$ (which corresponds to neglecting the small correction
$\asini\ll\Dt\sim T$), but numerically we have $\fbar=f$ in the present constant-frequency
case without spindowns.
\newlength{\lenminus}
\settowidth{\lenminus}{$-$}
\newcommand{\lm}{\hspace*{\lenminus}}
Hence, we obtain the phase derivatives
\begin{align}
  \frac{\partial_\freq \Phase}{2\pi}  &\approx \lm\Dt\,,  \notag \\
  \frac{\partial_{\asini}\Phase}{2\pi} &= -\fbar \left[ \sin\psi + \frac{\kappa}{2}\sin2\psi - \frac{\eta}{2}\cos2\psi \right]\,,\notag\\
  \frac{\partial_{\tAsc}\Phase}{2\pi}  &= \lm\fbar\asini\Om\left[\cos\psi+\kappa\cos2\psi+\eta\sin2\psi\right]\,,\label{eq:78}\\
  \frac{\partial_{\Om}\Phase}{2\pi}   &= -\fbar\asini(t-\tAsc)[\cos\psi+\kappa\cos2\psi+\eta\sin2\psi],\notag\\
  \frac{\partial_{\kappa}\Phase}{2\pi} &= -\frac{1}{2}\fbar\asini\sin2\psi\,,\notag\\
  \frac{\partial_{\eta}\Phase}{2\pi}  &= \lm \frac{1}{2}\fbar\asini\cos2\psi\,, \notag
\end{align}
which are inserted into Eq.~\eqref{eq:7} in order to obtain the coherent metric. The semi-coherent metric is
obtained by averaging the coherent metrics over segments according to Eq.~\eqref{eq:8}.

The resulting analytic expressions for these metrics \emph{in the general case} are quite
uninstructive and unwieldy, while it is straightforward to compute them numerically for
any case of interest.
However, as noticed in previous investigations \cite{Dhurandhar:2000sd,Messenger:2011rg},
it is instructive to focus on two limiting regimes that yield particularly simple
analytical results, namely the \emph{long-segment limit} ($\LongSeg$) where $\Tseg \gg P$, and the
\emph{short-segment limit} ($\ShortSeg$) where $\Tseg\ll P$.

When taking these limits on the metric $g_{ij}$, in order to decide whether a particular
off-diagonal term is negligible or not, it is useful to consider the \emph{diagonal-rescaled}
metric
\begin{equation}
  \label{eq:32}
  \DN{g}_{ij} \equiv \frac{g_{ij}}{\sqrt{g_{ii}\,g_{jj}}}\,,
\end{equation}
which is dimensionless and has unit diagonal, $\DN{g}_{ii}=1$.
In this rescaled metric we can then naturally neglect off-diagonal terms if they satisfy
$\DN{g}_{ij}\ll1$, as their corresponding contribution to the metric mismatch of Eq.~\eqref{eq:11}
will then be negligible.

\subsection{The long-segment (LS) regime ($\Tseg\gg P$)}
\label{sec:long-segment-limit}

\subsubsection{Coherent metric $\coh{g}^\LongSeg$}
\label{sec:coherent-LS-metric}

As noted in \cite{Messenger:2011rg}, it is convenient to use the discrete gauge freedom in the
choice of the orbital reference epoch $\tAsc$, as we can choose the time of ascending node during any orbit.
Therefore we can add any integer multiple $p$ of a period and redefine
\begin{equation}
  \label{eq:33}
  \tAsc' = \tAsc + p \frac{2\pi}{\Om}\,,\quad p \in \Zn{},
\end{equation}
without changing the system.
An (infinitesimal) \emph{offset} (or ``uncertainty'') $\delta\tAsc$ changes under such a
transformation in the presence of an (infinitesimal) offset on the period, i.e.\ if $\delta\Om\not=0$, namely
\begin{equation}
  \label{eq:34}
  \delta\tAsc' = \delta\tAsc - p\frac{2\pi}{\Om^2}\,\delta\Om\,.
\end{equation}
All other coordinate offsets are unaffected by this change of orbital reference epoch.
This needs to be carefully taken into account in the metric when redefining $\tAsc$.
Namely, given that this is a pure ``relabeling'' of the same physical situation,
the corresponding metric mismatch must be invariant, i.e.\
\begin{align}
  \label{eq:35}
  \mis =& \delta\Dop^{'m}\,g'_{m b}\,\delta\Dop^{'b} \notag\\
       =& \delta\Dop^{i} \left[\frac{\partial\Dop^{'m}}{\partial\Dop^{i}} g'_{m b} \frac{\partial\Dop^{'b}}{\partial\Dop^{j}}\right]\,\delta\Dop^{j} \notag\\
       =& \delta\Dop^i \, g_{ij}\,\delta\Dop^j\,,
\end{align}
where the only nonzero components of the Jacobian of this coordinate transformation
Eq.~\eqref{eq:33} are
\begin{equation}
  \label{eq:36}
  \frac{\partial\Dop^{'i}}{\partial\Dop^i} = 1\,,\quad\text{and}\quad
  \frac{\partial\tAsc'}{\partial\Om} = -p\,\frac{2\pi}{\Om^2}\,.
\end{equation}
In the long-segment limit, these discrete steps $P=2\pi/\Om$ are assumed to be small compared to the
segment length $\Tseg$. We can therefore choose the time of ascending node to be approximately at the
\emph{segment midtime} $\tMid$, i.e. we consider the special gauge choice $\tAsc' \approx \tMid$, in
which the phase derivatives Eq.~\eqref{eq:78} are time-symmetric around $\tMid$,
which simplifies the metric calculation.
Introducing the time offset
\begin{equation}
  \label{eq:38}
  \Dma \equiv \tMid - \tAsc\,,
\end{equation}
we therefor consider first the special case $\coh{g}^\LongSeg_{\atmid,ij}\equiv \coh{g}_{ij}^\LongSeg(\Dma=0)$.

Keeping only leading-order terms in $\Om\,\Tseg\gg2\pi$, and up to first order in $\ecc$
(i.e.\ first order in $\kappa$ and $\eta$), we obtain the diagonal metric
\begin{align}
  \label{eq:37}
  &\coh{g}^\LongSeg_{\atmid,ii} = 2\pi^2\,\times\\
  &\left[
    \frac{\Tseg^2}{6},\,{\fbar}^2,\,(\fbar\asini\Om)^2,\,\frac{(\fbar\asini\Tseg)^2}{12},\frac{(\fbar\asini)^2}{4},\frac{(\fbar\asini)^2}{4}\nonumber
  \right]\,,
\end{align}
in agreement with the result Eq.~(30) of \cite{Messenger:2011rg}.

The general case of $\Dma\not=0$ can be obtained by applying the Jacobian\footnote{A direct but
  somewhat more complicated calculation of the metric at $\Dma\not=0$ yields the same result}
of Eq.~\eqref{eq:36}
with $\Dma = p\,2\pi/\Om$, and so $\tAsc' = \tAsc + \Dma$ and
$\partial\tAsc'/\partial\Om = -\Dma/\Om$, resulting in the \emph{general} coherent long-segment metric
$\coh{g}^\LongSeg_{ij}(\Dma)$ with components
\begin{align}
  \coh{g}^\LongSeg_{\freq\freq}  &= \pi^2\, \frac{\Tseg^2}{3}\,,\notag\\
  \coh{g}^\LongSeg_{\asini\asini} &= 2\pi^2\, \fbar^2\,, \notag\\
  \coh{g}^\LongSeg_{\tAsc\tAsc}  &= 2\pi^2\, (\fbar\asini\Om)^2\,,\label{eq:39}\\
  \coh{g}^\LongSeg_{\Om\Om}     &= 2\pi^2\, (\fbar\asini)^2\,\left(\frac{\Tseg^2}{12} + \Dma^2\right)\,,\notag\\
  \coh{g}^\LongSeg_{\Om\tAsc}  &= \coh{g}^\LongSeg_{\tAsc\Om} = -2\pi^2\, (\fbar\asini)^2\,\Om\,\Dma\,,\notag\\
  \coh{g}^\LongSeg_{\kappa\kappa} &= \coh{g}^\LongSeg_{\eta\eta} = \frac{\pi^2}{2}\,(\fbar\asini)^2,\notag
\end{align}
while all other off-diagonal terms are (approximately) zero in this limit.
The nonzero off-diagonal term shows that in general (i.e.\ for $\Dma\not=0$) there are correlations between
offsets in $\tAsc$ and $\Om$, introduced by the relation Eq.~\eqref{eq:34}.
This generalizes the result in \cite{Messenger:2011rg} to arbitrary choices of orbital reference
epoch $\tAsc$.

\subsubsection{Semi-coherent metric $\inc{g}^\LongSeg$}
\label{sec:semi-coherent-LS-metric}

In order to compute the semi-coherent metric, we only need to average the per-segment coherent
metrics of the previous sections over segments according to Eq.~\eqref{eq:8}. Here it is crucial to
realize that all segment metrics must use the \emph{same coordinates} $\Dop$ in order for
this averaging expression to apply, in particular we can only fix the gauge on $\tAsc$ \emph{once}
for all segments, and so generally $\Dmal \not=0$. This means we \emph{cannot} use the special form
of $\coh{g}^\LongSeg_{\atmid}$ of Eq.~\eqref{eq:37} for the average over all segments, as done in
\cite{Messenger:2011rg}, which would result in the erroneous conclusion that the semi-coherent metric
$\inc{g}^\LongSeg$ would be identical to the coherent one.
Instead, using Eq.~\eqref{eq:8}, we find by averaging the per-segment metrics
$\coh{g}_\iSeg^\LongSeg \equiv \coh{g}^\LongSeg(\Dmal)$:
\begin{align}
   \inc{g}^\LongSeg_{\freq\freq}  &= \pi^2\, \frac{\Tseg^2}{3}\,,\notag\\
  \inc{g}^\LongSeg_{\asini\asini}&= 2\pi^2\, \fbar^2\,, \notag\\
  \inc{g}^\LongSeg_{\tAsc\tAsc} &= 2\pi^2\, (\fbar\asini\Om)^2\,,   \label{eq:40}\\
  \inc{g}^\LongSeg_{\Om\Om}    &= 2\pi^2\, (\fbar\asini)^2\left(\frac{\Tseg^2}{12} + \avgN{\Dma^2}\right)\,,\notag\\
  \inc{g}^\LongSeg_{\Om\tAsc}  &= \inc{g}^\LongSeg_{\tAsc\Om} = -2\pi^2\, (\fbar\asini)^2\,\Om\,\avgN{\Dma}\,, \notag\\
  \inc{g}^\LongSeg_{\kappa\kappa}&= \inc{g}^\LongSeg_{\eta\eta} = \frac{\pi^2}{2}\,(\fbar\asini)^2\,  \notag
\end{align}
It is interesting to consider the ideal case of $\Nseg$ regularly-spaced segments without gaps,
i.e.\ $\Tobs = \Nseg\,\Tseg$, and
\begin{equation}
  \label{eq:68}
  \Dmal = (\avgN{\tMid} - \tAsc) + \left(\iSeg - \frac{\Nseg+1}{2}\right)\,\Tseg\,,
\end{equation}
resulting in the first two moments
\begin{align}
  \label{eq:80}
  \avgN{\Dma} &= \avgN{\tMid} - \tAsc\,,\\
  \avgN{\Dma^2} &= \avgN{\Dma}^2 + \frac{1}{12}(\Nseg^2 - 1)\Tseg^2\,,
\end{align}
where $\avgN{\tMid}$ is the average mid-time over all segments.
In this case we can write the variance of $\{\Dmal\}$ as
\begin{equation}
  \label{eq:69}
  \var{\Dma} \equiv \avgN{\Dma^2} - \avgN{\Dma}^2 = \frac{1}{12}(\Nseg^2 - 1)\Tseg^2\,.
\end{equation}
If we use the gauge freedom and choose $\tAsc\approx\avgN{\tMid}$, then $\avgN{\Dma} \approx 0$,
and $\avgN{\Dma^2} \approx (\Nseg^2-1)\,\Tseg^2/12$, so that in this case the metric is again diagonal, and
the only component different from the coherent metric $\coh{g}^\LongSeg_{\atmid}$ is
\begin{equation}
  \label{eq:41}
  \inc{g}^\LongSeg_{\Om\Om} = \pi^2\, \frac{(\fbar\asini)^2}{6}(\Nseg\,\Tseg)^2 \propto \Tobs^2\,,
\end{equation}
corresponding to a refinement in the $\Om$ coordinate.
Therefore the semi-coherent template-bank spacing in $\Om$ needs to be finer by a factor of $\Nseg$
compared to coherent spacing in order to achieve the same mismatch.
The existence of this metric refinement in $\Om$ in the long-segment limit had already been noticed
earlier~\cite{holger:_refinement}, and was also used in the discovery of a binary pulsar in Fermi LAT
data~\cite{2012Sci...338.1314P}.

\subsection{The short-segment (SS) regime ($\Tseg\ll P$)}
\label{sec:short-segment-limit}

\subsubsection{Coherent metric $\coh{g}^\ShortSeg$}
\label{sec:coherent-SS-metric}

In the case of very short coherent segments compared to the period, i.e.\ $\Om\Tseg\ll2\pi$, as
pointed out in \cite{Messenger:2011rg}, strong parameter-space degeneracies render
the metric in these coordinates quite impractical for template-bank generation, and we therefore
follow their approach of Taylor-expanding the phase around the mid-point $\tMid$ of the coherent
observation time, namely
\begin{equation}
  \label{eq:42}
  \Phase(t) = 2\pi \,\sum_{k=1}^{n} \frac{u_k}{k!} (t - \tMid)^k\,,
\end{equation}
omitting a constant term.

The $\uCoord$-coordinates $\{u_k\}$ are defined as the k-th time-derivatives of the phase at $\tMid$, namely
\begin{equation}
  \label{eq:43}
  u_k \equiv \frac{1}{2\pi} \left. \partial_t^k \Phase(t) \right|_{\tMid}\,.
\end{equation}
Note that the phase model of Eq.~\eqref{eq:42} is formally identical to the phase in terms of
the frequency and spindowns $\{\freq^{(k)}\}$ of an isolated NS (with $\tMid$ playing the role of the reference time
$\tRef$), as can be seen in Eq.~\eqref{eq:23}. The corresponding ``spindown metric'' has a
well-known analytical form, which can be expressed most conveniently in terms of the rescaled
dimensionless $v$-coordinates,
\begin{equation}
  \label{eq:45}
  {v}_k \equiv 2\pi\, \frac{u_k}{k!}\left(\frac{\Tseg}{2}\right)^k\,,
\end{equation}
resulting in the $v$-coordinate metric
\begin{align}
  \label{eq:46}
  \coh{g}^{\ShortSeg,{v}}_{k k'}&= \frac{\modulo{(k + k' - 1)}{2}}{k + k' + 1} \notag\\
  & \hspace{1cm}- \frac{[\modulo{(k - 1)}{2}]\,\,[\modulo{(k' - 1)}{2}]}{(k+1)(k'+1)}\notag\\
    &= \begin{pmatrix}
    \frac{1}{3}   &            0  &  \frac{1}{5}  &             0  & \ldots \\[0.1cm]
          0       & \frac{4}{45}  &             0 & \frac{8}{105}  & \ldots \\[0.1cm]
     \frac{1}{5}  &          0    & \frac{1}{7}   &             0  & \ldots \\[0.1cm]
          0       & \frac{8}{105} &       0       & \frac{16}{225} & \ldots \\
          \vdots  &    \vdots     &     \vdots    &    \vdots      & \ddots
  \end{pmatrix}\,,
\end{align}
correcting the incorrect expression Eq.~(21) of \cite{Messenger:2011rg}.

Applying Eq.~\eqref{eq:43} to the small-eccentricity phase of Eq.~\eqref{eq:30}, we obtain the
following expressions
\begin{align}
u_1 &=   -\fbar\asini\Om\,[\cos\psim +  \kappa\cos2\psim + \eta\sin2\psim ] + \freq, \notag\\
u_2 &= \lm\fbar\asini\Om^2\,[\sin\psim + 2\kappa\sin2\psim - 2\eta\cos2\psim ],\notag\\
u_3 &= \lm\fbar\asini\Om^3\,[\cos\psim + 4\kappa\cos2\psim + 4\eta\sin2\psim ],\label{eq:ucoords}\\
u_4 &=   -\fbar\asini\Om^4\,[\sin\psim + 8\kappa\sin2\psim - 8\eta\cos2\psim ],\notag\\
u_5 &=   -\fbar\asini\Om^5[\cos\psim +16\kappa\cos\!2\psim + 16\eta\sin2\psim],\notag\\
u_6 &= \lm\fbar\asini\Om^6[\sin\psim +32\kappa\sin2\psim - 32\eta\cos2\psim ],\notag\\
\vdots\nonumber
\end{align}
with $\psim \equiv \psi(\tMid)$, which differ in the sign of $\asini$ compared to Eqs.~(A2-A5) of
\cite{Messenger:2011rg}, due to the sign error on $R$ in the phase model discussed in
Sec.~\ref{sec:gener-newt-binary}.

Note that for the general elliptical case we expect $6$ independent $\uCoord$-coordinates $\uCoord(\Dop)$
and $4$ in the circular case.
The nonlinear coordinate transformation Eqs.~\eqref{eq:ucoords} can
be analytically inverted to obtain $\Dop(u)$, as shown explicitly in Appendix~\ref{sec:invert-u-coord}.

We follow \cite{Messenger:2011rg} in estimating the range of validity of this Taylor approximation
up to order $n$ by considering the order of magnitude of the first neglected term of order $n+1$,
namely
\begin{align}
  \label{eq:62}
  |\Delta\Phase| &\sim 2\pi \frac{u_{n+1}}{(n+1)!}\left(\frac{\Tseg}{2}\right)^{n+1}\,,\notag\\
  &\sim 2\pi \,\fbar\asini \frac{(\pi\,x)^{n+1}}{(n+1)!}\,,
\end{align}
where we defined the fraction $x$ of an orbit sweeped during the segment duration, i.e.\
$x \equiv \Om\Tseg/(2\pi) = \Tseg/P$.
In order to link this phase error to a mismatch, we observe that here $\Delta\Phase\sim \Delta v_{n+1}$ of
Eq.~\eqref{eq:45}, and the corresponding metric element of
Eq.~\eqref{eq:46} gives  $\coh{g}^{\ShortSeg,v}_{(n+1),(n+1)} \sim \frac{1}{2n+3}$ for $n \gg 1$ and therefore
\begin{equation}
  \label{eq:63}
  \mis \sim \frac{\Delta\Phase^2}{2n+3} = (2\fbar\asini)^2 \frac{(\pi\,x)^{2(n+1)}}{(2n+3)\,{(n+1)!}^{2}}\,.
\end{equation}
Plugging in typically values used in the numerical tests later, e.g.\
$\fbar\sim 500\,$Hz, $\asini\sim3\,$s, $n=6$, we see that the mismatch grows very rapidly from
$\mis\ll1$ at $x\lesssim0.25$ to $\mis\sim1$ at $x\sim0.35$. We therefore expect this approximation
with $n=6$ to typically hold for segment durations up to a $1/4$ to $1/3$ of an orbit.

\subsubsection{Semi-coherent metric $\inc{g}^\ShortSeg$}
\label{sec:semi-coherent-SS-metric}

While the segments $\Tseg$ are assumed to be short compared to the period $P$, we only consider
the case where the total observation time is long compared to $P$ i.e.\ in addition to
$\Tseg\ll P$ we also assume $\Tobs = \Nseg\Tseg\gg P$, which also implies $\Nseg\gg 1$.

We cannot directly use the coherent per-segment metrics in $\uCoord$-coordinates to compute the
semi-coherent metric as an average using Eq.~\eqref{eq:8}, because a fixed physical parameter-space
point $\Dop$ would have different $\uCoord$-coordinates in each segment.

Instead, as first shown in \cite{Messenger:2011rg}, one can go back to physical coordinates and use
the fact that $\Nseg\gg1$ to replace the discrete sum in Eq.~\eqref{eq:8} over segments by an integral over time, namely
\begin{align}
  \label{eq:47}
  \inc{g}^\ShortSeg &= \frac{1}{\Nseg}\sum_{\ell=1}^\Nseg \coh{g}^\ShortSeg(\tMidl) \notag\\
  &\stackrel{\Nseg\gg1}{\approx} \frac{1}{\Tobs}\int_{\tMid-\Tobs/2}^{\tMid+\Tobs/2}\coh{g}^\ShortSeg(t')\,dt'\,,
\end{align}
in terms of coherent per-segment metrics $\coh{g}^{\ShortSeg}(\tMidl)$ expressed as a function of the
segment mid-times $\tMidl$, and where $\tMid$ is the mid-time of the whole observation.
In order to compute this, we first calculate the coherent metric of Eq.~\eqref{eq:7} for each
segment $[\tMidl-\Tseg/2,\,\tMidl+\Tseg/2]$ by using the phase derivatives of Eq.~\eqref{eq:78}.
As before, we keep only first-order terms in $\ecc$ (i.e.\ first-order in $\kappa,\eta$), then
Taylor-expand the results in $\Om\,\Tseg\ll2\pi$ up to second order, obtaining $\coh{g}^{\ShortSeg}(\tMidl)$.
This is integrated over the total observation time according to Eq.~\eqref{eq:47}, and only
leading-order terms in $\Om\,\Tobs\gg 2\pi$ are kept. We only keep off-diagonal terms where the
corresponding diagonal-rescaled elements Eq.~\eqref{eq:32} are not $\ll 1$.
The resulting semi-coherent short-segment long-observation limit metric elements are found as
\begin{align}
  \inc{g}^\ShortSeg_{\freq\freq}  &= \pi^2\, \frac{\Tseg^2}{3}\,,   \notag \\
  \inc{g}^\ShortSeg_{\asini\asini}&= \frac{\pi^2}{6}(\Om\Tseg)^2\, \fbar^2\,,\notag\\
  \inc{g}^\ShortSeg_{\tAsc\tAsc} &=  \frac{\pi^2}{6}(\Om\Tseg)^2\,(\fbar\asini\Om)^2\,,\label{eq:48}\\
  \inc{g}^\ShortSeg_{\Om\Om}    &=  \frac{\pi^2}{6}(\Om\Tseg)^2\,(\fbar\asini)^2\left(\frac{\Tobs^2}{12}+\avgN{\Dma}^2\right)\,,\notag\\
  \inc{g}^\ShortSeg_{\Om\tAsc}  &= \inc{g}^\ShortSeg_{\tAsc\Om} = -\frac{\pi^2}{6}(\Om\Tseg)^2\,(\fbar\asini)^2\,\Om\,\avgN{\Dma}\,.\notag\\
  \inc{g}^\ShortSeg_{\kappa\kappa}&= \inc{g}^\ShortSeg_{\eta\eta} = \frac{\pi^2}{6}(\Om\Tseg)^2\,(\fbar\asini)^2\,,\notag
\end{align}
which generalizes the result in \cite{Messenger:2011rg} (for gauge choice $\tAsc\approx\avgN{\tMid}$, i.e.\
$\avgN{\Dma}\approx0$) to the general case of $\avgN{\Dma}\not=0$.

Note that a more general form encompassing the semi-coherent metric in both the short-segment and
long-segment limits has recently found in \cite{whelan2015:_cross_corr}.

\section{Number of templates}
\label{sec:number-templates}

In order to express the explicit template counts based on Eq.~\eqref{eq:NumTemplates}, we will
assume a simple parameter space $\DopSpace$ bounded by $\fbar\in[\Min{\fbar},\Max{\fbar}]$,
$\asini\in[\Min{\asini},\Max{\asini}]$, $\tAsc\in[\Min{\tAsc},\Max{\tAsc}]$,
$\Om\in[\Min{\Om},\Max{\Om}]$, $\ecc\in[\Min{\ecc},\Max{\ecc}]$, and
$\argp\in[\Min{\argp},\Max{\argp}]$.
Using Eq.~\eqref{eq:26}, \eqref{eq:26b} we see that
\begin{equation}
  \label{eq:9}
  \int\int \,d\kappa\,d\eta = \int\int\,\ecc\,d\ecc\,d\argp\,.
\end{equation}
Observing that $g_{\kappa\kappa}=g_{\eta\eta}$, we can further obtain
$\mis = g_{\kappa\kappa}\,d\kappa^2+g_{\eta\eta}\,d\eta^2 = g_{\kappa\kappa}\,(d\ecc^2 +
\ecc^2\,d\argp^2)$, which implies
\begin{equation}
  \label{eq:83}
  g_{\ecc\ecc} = g_{\kappa\kappa}\,,\quad
  g_{\argp\argp} = \ecc^2\,g_{\kappa\kappa}\,,\quad
  g_{\ecc\argp} = 0\,.
\end{equation}
For convenience of notation, for the following template-count expressions, we define the shorthand
\begin{equation}
  \label{eq:64}
  \MaxMin{Q} \equiv \Max{Q} - \Min{Q}\,.
\end{equation}

\subsection{Templates $\Nt^\LongSeg$ in the long-segment regime}
\label{sec:num-templates-long-segm}

The determinant of the semi-coherent metric $\inc{g}^\LongSeg$ of Eq.~\eqref{eq:40} only
differs from that of the coherent metric $\coh{g}^\LongSeg$ of Eq.~\eqref{eq:39} by the
presence of a refinement factor $\refine$, i.e.
\begin{align}
  \label{eq:84}
  \det\inc{g}^\LongSeg &= \refine^2\;\det\coh{g}^\LongSeg\,,\quad\text{with}\\
  \refine &\equiv \sqrt{1 + 12\frac{\var{\Dma}}{\Tseg^2}} \stackrel{\text{gapless}}{=} \Nseg\,,
\end{align}
where the last equality holds in the special case of $\Nseg$ segments without gaps, as seen from
Eq.~\eqref{eq:69}. This refinement only affects the $\Om$-dimension. In particular, for the gauge
choice $\avgN{\Dma}=0$, we see
\begin{equation}
  \label{eq:91}
  \inc{g}^\LongSeg_{\atmid,\Om\Om} = \refine^2\,\coh{g}^\LongSeg_{\atmid,\Om\Om}\,.
\end{equation}
The determinant Eq.~\eqref{eq:84} is seen to be independent of the gauge choice on $\tAsc$ (i.e.\ on $\avgN{\Dma}$), and we
can therefore write the template volume density in factored form as
\begin{equation}
  \label{eq:6}
  \sqrt{\det g^\LongSeg}\,d^n\Dop = \prod_{i=1}^n \sqrt{g^\LongSeg_{\atmid,ii}}\,d\Dop^i \,.
\end{equation}
This independence of $\Dma$ in the volume element may seem surprising at first, as one would expect
\cite{2014ApJ...781...14G} the parameter-space uncertainty on $\tAsc$ to grow with increasing
separation from the observation time $\tMid = \tAsc + \Dma$, if there is any uncertainty $d\Om$ in
the mean angular velocity, as also discussed in Sec.~\ref{sec:long-segment-limit}.
However, this is only the case if the uncertainty in $\Om$ is not resolved by the template bank,
when one observes indeed a stretching of the effective (projected) uncertainty in $\tAsc$,
resulting in an increase in required templates $\Nt_{\tAsc}$.
On the other hand, if the $\Om-\tAsc$ parameter space is fully resolved by the metric [as
assumed in Eq.~\eqref{eq:6}], there is no increase of templates.
Changing $\Dma$ only deforms the $\Om-\tAsc$ parameter space in a volume-preserving way by
shear according to Eq.~\eqref{eq:33}. Given that any offset in $\tAsc$ and $\Om$ results in a
deterministically shifted offset $\tAsc'$, there is no loss of information, and
therefore no increase in the number of templates.
Alternatively, one can consider the orbital epoch $\tAsc$ with its uncertainty fixed at its original
measurement epoch, and account for the offset $\Dma$ in the metric as done in
Sec.~\ref{sec:long-segment-limit}.
This results in changing metric correlations between $\tAsc$ and $\Om$ while leaving the ellipse
volume unchanged.

The expression for the number of templates per dimension $\NtPerDim{\Dop^i}$
of Eqs.~\eqref{eq:82} and \eqref{eq:89} simplifies to
\begin{equation}
  \label{eq:90}
  \NtPerDim{\Dop^i}^\LongSeg = \frac{1}{2}\,\misMax^{-1/2}\,\sqrt{g_{\atmid,ii}^\LongSeg}\;\Delta\Dop^i\,,
\end{equation}
with individual components in coordinates $\Dop^i = \{\freq,\asini,\tAsc,\Om,\ecc\}$ given by
\begin{equation}
  \label{eq:92}
  \sqrt{g_{\atmid,ii}^\LongSeg} = \pi\sqrt{2} \left[
    \frac{\Tseg}{\sqrt{6}}\,,
    \fbar,\,
    \fbar\asini\Om,\,
    \fbar\asini\,\frac{\refine\Tseg}{\sqrt{12}},\,
    \frac{\fbar\asini}{2}\,
  \right]\,,
\end{equation}
noting that $g_{\ecc\ecc} = g_{\kappa\kappa} = g_{\eta\eta}$ and
$g_{\argp\argp}=\ecc^2\,g_{\ecc\ecc}$.
In practice, one often encounters cases where the parameter uncertainty along some of these
dimensions is smaller than the metric resolution, such that a single template covers the whole
extent of the parameter space along that direction.
In this case the corresponding coordinate contribution to the template density would result in
fractional templates (and therefore underestimating the number of templates) and must not be
included in Eq.~\eqref{eq:NumTemplates}, as discussed in Sec.~\ref{sec:templ-banks-metr}.

The number of templates over the full 6D parameter space $\{\freq,\asini,\tAsc,\Om,\ecc,\argp\}$
according to Eq.~\eqref{eq:NumTemplates} is obtained as
\begin{equation}
  \label{eq:31N}
  \Nt^\LongSeg = \frac{\thick{6}}{\misMax^{3}}\,\frac{\pi^6\,\refine\Tseg^2}{360\sqrt{2}}\,
  \MaxMin{\fbar^6}\MaxMin{\asini^5}\MaxMin{\tAsc}\MaxMin{\Om^2}
  \MaxMin{\ecc^2}\MaxMin{\argp}\,.
\end{equation}
In the application to Scorpius~X-1, a few special cases will be of interest, namely when one or more of
the uncertainties in orbital parameters are smaller than the template extent.

For sufficiently well-estimated orbital angular velocity $\Om=\Om_0$ (such that $\NtPerDim{\Om}<1$), the
template count for the corresponding 5D template bank over $\{\freq,\asini,\tAsc,\ecc,\argp\}$ is found as
\begin{equation}
  \label{eq:85}
\Nt^\LongSeg = \frac{\thick{5}}{\misMax^{5/2}}
  \frac{\pi^5\,\Tseg}{40\,\sqrt{3}}\,\Om_0\,\MaxMin{\fbar^5}\MaxMin{\asini^4}\MaxMin{\tAsc}\MaxMin{\ecc^2}\MaxMin{\argp}\,,
\end{equation}
while in the 4D case of well-estimated $\ecc$ and $\argp$ (or equivalently, for circular orbits)
$\{\freq,\asini,\tAsc,\Om\}$, we obtain
\begin{equation}
  \label{eq:66}
  \Nt^\LongSeg = \frac{\thick{4}}{\misMax^2}\,\frac{\pi^4\,\refine\Tseg^2}{36\,\sqrt{2}}\,\MaxMin{\fbar^4}\MaxMin{\asini^3}\MaxMin{\tAsc}\MaxMin{\Om^2}\,,
\end{equation}
and finally in the 3D circular case with well-determined $\Om=\Om_0$, $\ecc$ and $\argp$ we find the
template count over the remaining parameter space $\{\freq,\asini,\tAsc\}$ as
\begin{equation}
  \label{eq:86}
  \Nt^\LongSeg =
  \frac{\thick{3}}{\misMax^{3/2}}\frac{\pi^3\,\Tseg}{\sqrt{27}}\,\Om_0\,
  \MaxMin{\fbar^3}\MaxMin{\asini^2}\MaxMin{\tAsc}\,.
\end{equation}
We note that Eqs.~\eqref{eq:92}--\eqref{eq:86} are valid for the semi-coherent case with general refinement $\refine$,
as well as for the coherent case with $\refine=1$.

\subsection{Templates $\Nt^\ShortSeg$ in the short-segment regime}
\label{sec:num-templates-short-segm}

Due to the nonlinear transformation  Eqs.~\eqref{eq:ucoords} from physical
parameters $\Dop$ into $\uCoord$-coordinates, the physical parameter space $\DopSpace$ would be
described by complicated integration boundaries in $\uCoord$.
In addition, the number of dimensions in $\uCoord$-coordinates that need to be included in the
template bank is a non-trivial function of the signal parameters and the mismatch $\misMax$ (e.g.\ as
seen later in Fig.~\ref{fig:uOccupancy}).
These effects would dominate the expression for the number of templates Eq.~\eqref{eq:NumTemplates}, and it
is therefore not clear whether a useful closed analytical form can be given.

Using the metric volume density for the semi-coherent short-segment metric $\inc{g}^\ShortSeg$ of
Eq.~\eqref{eq:48}, we obtain the following number of templates Eq.~\eqref{eq:NumTemplates}
for the full 6D template bank
\begin{equation}
  \label{eq:73}
  \inc{\Nt}^\ShortSeg = \frac{\thick{6}}{\misMax^{3}}\,\frac{\pi^6\,\Tseg^6\,\Tobs}{90720\sqrt{6}}\,
  \MaxMin{\fbar^6}\MaxMin{\asini^5}\MaxMin{\tAsc}\MaxMin{\Om^7}
  \MaxMin{\ecc^2}\MaxMin{\argp},
\end{equation}
while in the 4D circular case we find
\begin{equation}
  \label{eq:75}
  \inc{\Nt}^\ShortSeg = \frac{\thick{4}}{\misMax^2}\,\frac{\pi^4\,\Tseg^4\Tobs}{2160\,\sqrt{6}}
  \,\MaxMin{\fbar^4}\MaxMin{\asini^3}\MaxMin{\tAsc}\MaxMin{\Om^5}\,.
\end{equation}

\section{Numerical Tests of the Metrics}
\label{softwaresim}

In this section we present numerical tests performed on the parameter-space metrics derived in the
previous sections. These tests consist in comparing the predicted metric mismatches $\mis$ of
Eq.~\eqref{eq:6b} against measured $\Fstat$-statistic mismatches $\misF$ of Eq.~\eqref{eq:5}
obtained via signal software injections.
The quantity used for this comparison is the (symmetric) relative
difference~\cite{Prix:2006wm,Wette:2013wza,Wette:2014a}, defined as
\begin{equation}\label{eq:RelErr}
  \relerr(\misF, \mis) \equiv \frac{\misF - \mis}{0.5\,(\misF + \mis)},
\end{equation}
which is bounded within $\relerr \in [-2, 2]$. For small values $|\relerr| \ll 1$ this agrees with
the asymmetric definitions of relative errors, such as $\relerr_0\equiv(\misF -\mis)/\misF$,
with domain $\relerr_0\in(-\infty,1]$, and which is related to the symmetric relative difference
$\relerr$ as $\relerr_0 = \frac{\relerr}{1+\relerr/2}$.

\subsection{Monte-Carlo software-injection method}
\label{sec:general-approach}

The general algorithm used for these software-injection tests is the following:
\begin{enumerate}
\item Pick random signal amplitude parameters $\Amp$ and phase-evolution parameters $\Dop_\sig$ from
  suitable priors.

\item Generate a phase-parameter offset $\delta\Dop = \Dop_\templ - \Dop_\sig$ by finding the
  closest lattice template $\Dop_\templ$ to the signal point $\Dop_\sig$.

\item Compute the phase metric $g_{ij}(\Dop_\sig)$ and metric mismatch
  $\mis(\Dop_\sig,\Dop_\templ)=g_{ij}\,\delta\Dop^i\delta\Dop^j$ according to Eq.~\eqref{eq:6b}.

\item Generate a (noise-free) data-set containing the signal $x = h(t;\Amp,\Dop_\sig)$.
  Compute the (coherent or semi-coherent) $\Fstat$-statistic at the injection point,
  $\Fstat(x;\Dop_\sig)$, and at the template, $\Fstat(x;\Dop_\templ)$.
  In noise-free data we have $\Fstat(x;\Dop) = \expect{\Fstat(x;\Dop)}$,
  therefore we can obtain $\rho^2(\Amp,\Dop_\sig;\Dop)$ via Eq.~\eqref{eq:3} or Eq.~\eqref{eq:4},
  respectively, which yields the measured mismatch $\misF$ via Eq.~\eqref{eq:5}.
  Injection and recovery is performed using the LALSuite software package~\cite{LALSuite}.
\end{enumerate}

This procedure is applied to test each of the four types of metric detailed in
Secs.~\ref{sec:long-segment-limit} and \ref{sec:short-segment-limit}.

\subsubsection*{Choice of signal parameters $\{\Amp,\,\Dop_\sig\}$}
\label{sec:1.-choice-signal}

The random signal amplitude parameters $\Amp$ are chosen as follows:
the scalar amplitude $\hO$ plays no role for the metric and is fixed to $\hO=1$.
The inclination angle is drawn from a uniform distribution in $\cosi\in[-1,1]$, the polarization angle within
$\psi\in[0,2\pi]$, and the (irrelevant) initial phase within $\phiO\in[0,2\pi]$.

The sky-position for all signals is fixed (without loss of generality) to that of Scorpius~X-1,
namely $(\alpha,\,\delta) = (4.276,-0.273)\,$rad, and we assume gapless data from the LIGO Hanford
detector (H)~\cite{AdvLIGORef:2010}.

The random signal phase-evolution parameters $\Dop_\sig\in\DopSpace$ are generated by drawing
them from uniform distributions over the ranges:
\begin{align}
  \freq &\in [50,\,1000]\,\Hz\,,   \notag\\
  \asini &\equiv \frac{a\,\sin{}i}{c} \in [1,\,5]\,\mathrm{s}, \notag\\
  P &\in [P_{0} - dP,\, P_{0} + dP]\,,\notag\\
  \tPeri &\in \left[\tMid-\frac{P}{2},\,\tMid + \frac{P}{2}\right]\,,\label{eq:31tPeri}\\
  \log_{10} \ecc &\in [-5,\,\log_{10}(0.9)]\,,\notag\\
  \argp &\in [0,\,2\pi]\,\mathrm{rad}\,.\notag
\end{align}
The orbital period ``scale'' $P_0$ is fixed for each set of software injections (specified later), while the
corresponding sampling range $dP = P_0^2\,d\Om/(2\pi)$ is given in terms of a range $d\Om$ in
orbital angular velocity, which is chosen as $d\Om \sim 1/\sqrt{g_{\Om\Om}}$, corresponding to a
roughly unity mismatch along $\Om$.
The motivation for this construction is twofold: on one hand we need to control the scale of the
orbital period in order to ensure the appropriate short-segment ($\Tseg\ll P$) or long-segment
($\Tseg\gg P$) limit is satisfied in the tests. On the other hand, we wish to randomize the period
over a range larger than the typical template-bank spacings in order to fully sample the
Wigner-Seitz cell of the template bank.

For the metric tests presented in this section we employ a Demodulation method of computing the
$\Fstat$-statistic, using Short Fourier Transforms (SFTs) of length
$\Tsft = 5$~s. Due to the linear phase model over each SFT employed by the search code, this
corresponds to a maximum error in phase of $ |\Delta\Phase| \sim 0.07\,$~rad over the investigated
parameter space (see Appendix~\ref{Sec:Tsft-choice} for more details).

\subsubsection*{Generating template-bank offsets $\delta\Dop$}
\label{sec:gener-offs-delt}

Sampling suitable offsets $\delta\Dop$ for metric testing has proved to be a subtle and difficult
point in previous metric studies \cite{Prix:2006wm,Wette:2013wza}.
Here we employ a recent innovation for generating ``natural'' offsets from a
\emph{virtual} template bank (e.g.\ see \cite{Wette:2014a} for more details). This method can be
applied whenever the metric $g_{ij}$ is constant (i.e.\ independent of $\Dop$), which allows for
constructing a lattice-based template bank for a given maximal mismatch $\misMax$ of
Eq.~\eqref{eq:11}, namely
\begin{equation}
  \label{eq:44}
  \misMax \ge \delta\Dop^i\,g_{ij}\,\delta\Dop^j = \delta\beta^l\,\delta_{lk}\,\delta\beta^k\,,
\end{equation}
where $\delta\beta^l\equiv \delta\Dop^i\,{A_i}^l$ in terms of the Cholesky
decomposition of the metric, i.e.\ $g_{ij} = {A_i}^l\,\delta_{lk}\,{A_j}^k$.
Further rescaling $\delta q^i \equiv (\cR/\sqrt{\misMax})\,\delta \beta^i$ yields the corresponding Euclidean covering equation
\begin{equation}
  \label{eq:50}
  \delta q^i\,\delta_{ij}\,\delta q^j \le \cR^2\,,
\end{equation}
in terms of the covering radius $\cR$. This requirement can be satisfied by
lattices~\cite{ConwaySloane99,Prix:2007ks} such as, for example, the simple $n$-dimensional hyper-cubic lattice
$\Zn{n}$ (with covering radius of $\cR_{\Zn{n}}=\sqrt{n}/2$), or the highly efficient covering lattice $\Ans{n}$.
For both $\Zn{n}$ and $\Ans{n}$ lattices, efficient algorithms exist for finding the closest lattice
point $q_\templ$, satisfying Eq.~\eqref{eq:50} for any given point $q_\sig\in \mathbb{R}^{n}$,
with $\delta q = q_\templ - q_\sig$. For example, for $\Zn{n}$ this is trivially given by component-wise rounding,
i.e.\ $q_\templ = \mathrm{round}[q_\sig]$.
We can use this construction to: \textit{(i)} easily find the closest lattice template $\Dop_\templ$ to
any given $\Dop_\sig$, \textit{(ii)} transform into lattice coordinates, i.e.\ $q_\sig = q(\Dop_\sig)$, (iii) find the
closest lattice point $q_\templ$ and invert back, i.e.\ $\Dop_\templ = \Dop(q_\templ)$.
The resulting distribution of offsets $\delta\Dop=\Dop_\templ - \Dop_\sig$ uniformly samples the
Wigner-Seitz cell of the corresponding lattice, which is the natural offset distribution for uniform
signal probability over the parameter space $\Dop_\sig\in\DopSpace$.

Note that while the coherent short-segment metric $\coh{g}^\ShortSeg$ of Eq.~\eqref{eq:46}
is \emph{explicitly flat} (i.e.\ constant), this is not
true for the other three cases, namely the semi-coherent short-segment metric of
Eq.~\eqref{eq:48}, the coherent long-segment metric of Eq.~\eqref{eq:39}, and the semi-coherent
long-segment metric of \eqref{eq:40}.
However, it is easy to see that diagonal-rescaling via Eq.~\eqref{eq:32} achieves explicit flatness
in all these three cases, namely by working in terms of rescaled coordinates
\begin{equation}
  \label{eq:LambPrime}
  \delta\Dop^{'\,i} \equiv \sqrt{g_{ii}}\, \delta\Dop^{i}.
\end{equation}
It might seem that the terms involving $\Dma\equiv\tMid - \tAsc$ depend on the coordinate
$\tAsc$, but this effect can be neglected in all these cases, as we always assume $\Tobs\gg P$, and
by choosing a gauge on $\tPeri$ as in Eq.~\eqref{eq:31tPeri}, we can effectively approximate
$\Dma/\Tobs\approx 0$ in the metric.

The following metric tests use a $\Zn{6}$ lattice and a maximum mismatch of $\misMaxInc=\misMaxCoh=0.3$.
This lattice is chosen purely for simplicity, and only serves to generate realistic signal offsets
for testing the metric. Using a different lattice (e.g.\ $\Ans{n}$) would slightly change the
distribution of sampled offsets, but would be inconsequential for the metric-testing results.
The maximal-mismatch value of $\misMax=0.3$ represents a fairly ``typical'' value for realistic
searches, but is still small enough so that measured $\Fstat$-statistic losses are only minimally
affected by higher-order corrections compared to the metric approximation (e.g.\ see
\cite{Prix:2006wm,Wette:2013wza}). Here we are mostly interested in the accuracy of the metric
within its range of applicability, while nonlinear deviations from the metric approximation would
warrant a separate study.

\subsection{Results in the long-segment regime}
\label{sec:results:LS}

\subsubsection{Coherent long-segment metric $\coh{g}^\LongSeg$}
\label{CohMetTcohGreatP}

For these tests we fix the ``scale'' of the orbital period to $P_0 = 19$~h (similar to Scorpius~X-1), and
vary the segment length $\Tseg$ from $3$~days to $39$~days in steps of 4 days.
We choose the parameter space range for the orbital velocity $\Om$ as two times the largest possible
metric spacing occurring over the parameter space, i.e.,
$d\Om = 2\,\max[(\coh{g}^{\LongSeg}_{\Om\Om})^{-1/2}] \sim 1.2 \times 10^{-7}$~s$^{-1}$.
The resulting parameter-space half-width in Period is $d P\sim 90\,$s, and therefore we have
at least $\Tseg/P \ge 3.8$.

\begin{figure*}[htbp]
  \raggedright \hspace*{0.5\columnwidth} (a)\\[-0.5cm]
  \hspace*{0.5\columnwidth}\includegraphics[clip,width=\columnwidth]{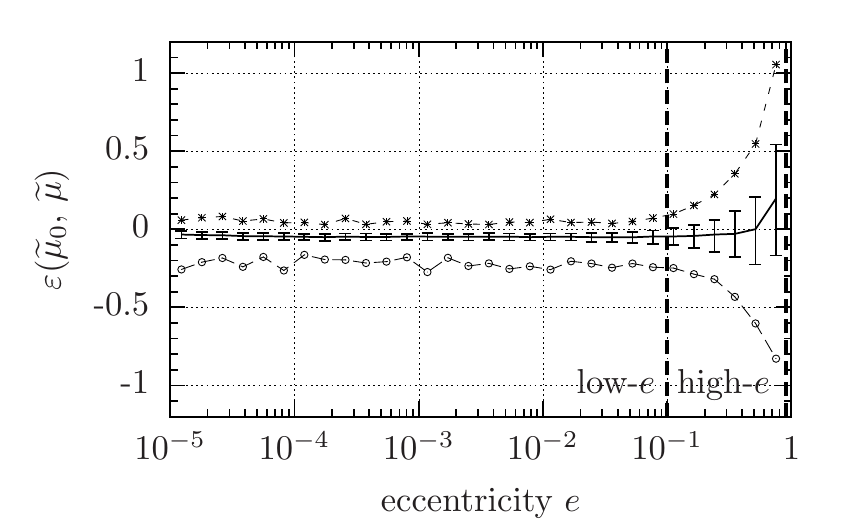}\\

  \raggedright (b)\hspace*{\columnwidth}(c)\\[-0.5cm]
  \includegraphics[clip,width=\columnwidth]{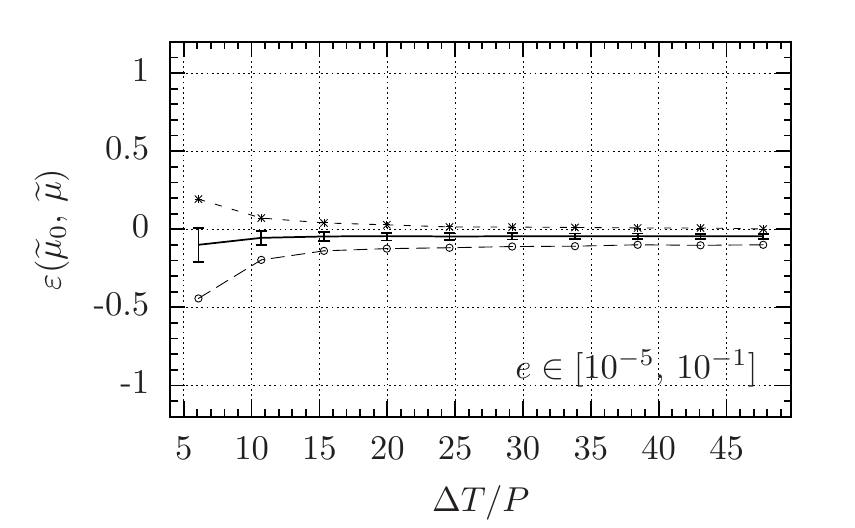}
  \includegraphics[clip,width=\columnwidth]{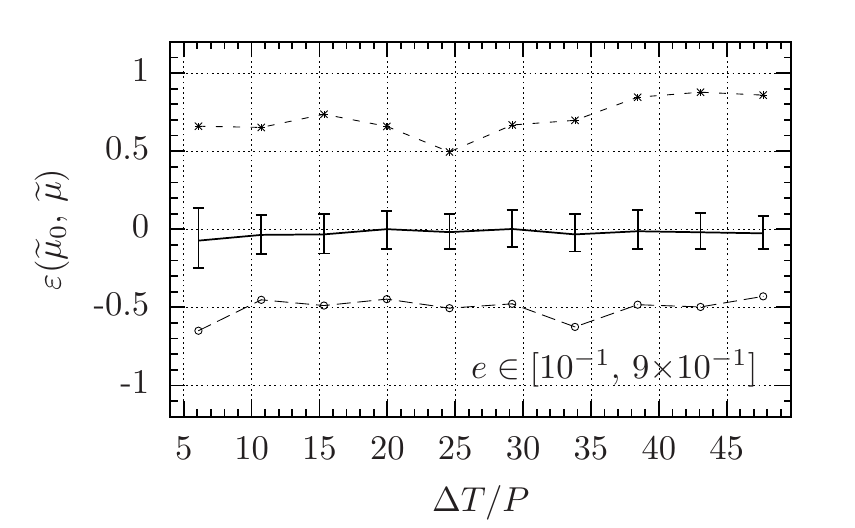}\\[-0.3cm]

  \raggedright (d)\hspace*{\columnwidth}(e)\\[-0.5cm]
  \includegraphics[clip,width=\columnwidth]{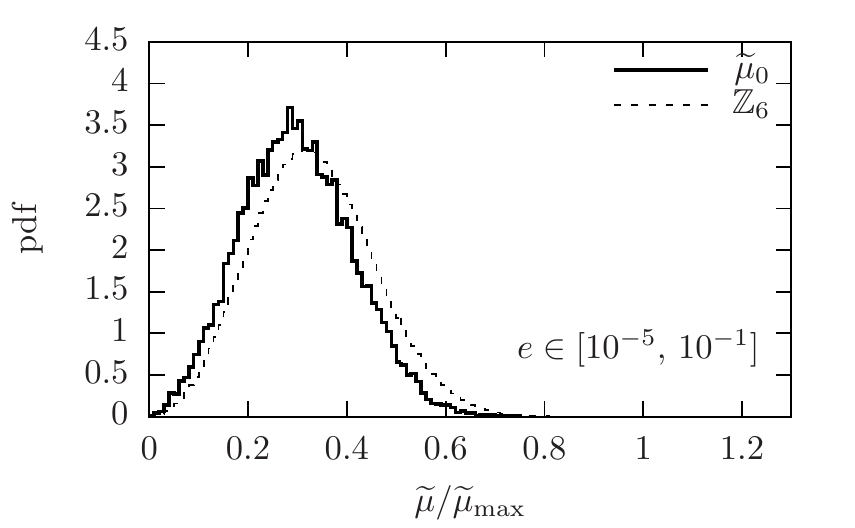}
  \includegraphics[clip,width=\columnwidth]{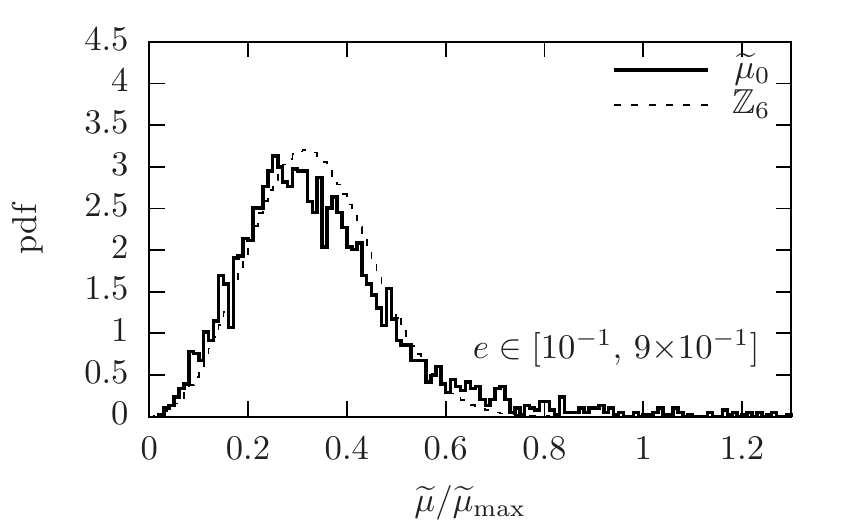}\\[-0.3cm]

  \raggedright (f)\hspace*{\columnwidth}(g)\\[-0.5cm]
  \includegraphics[clip,width=\columnwidth]{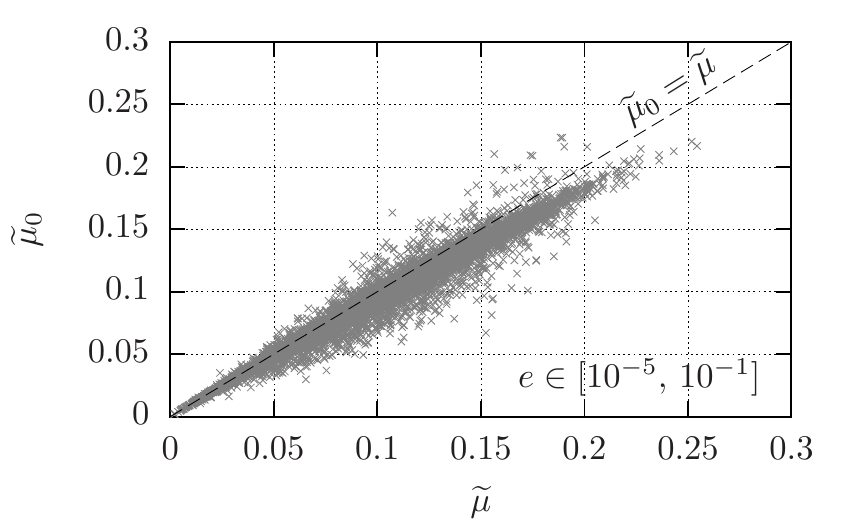}
  \includegraphics[clip,width=\columnwidth]{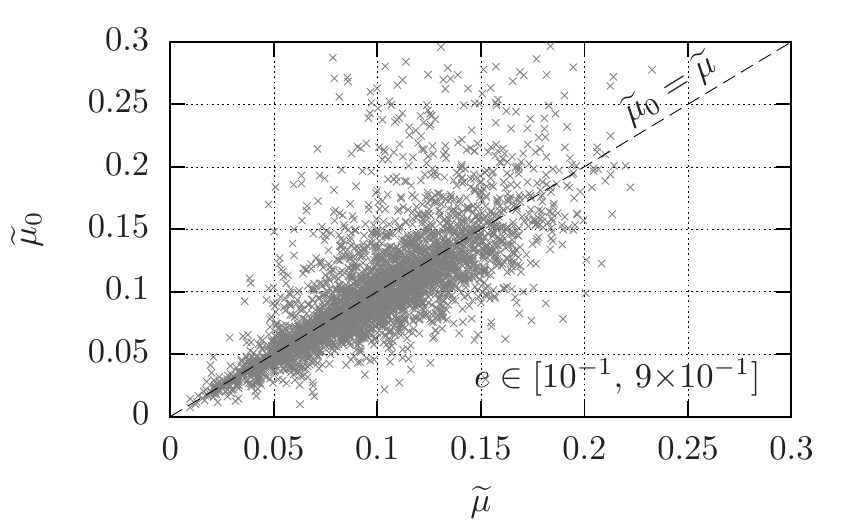}\\[-0.3cm]
  \caption{
    \emph{Coherent long-segment} regime: results for metric tests on $\coh{g}^\LongSeg$ of
    Sec.~\ref{sec:coherent-LS-metric}, for orbital period scale of $P_0 = 19$~h and varying segment
    length $\Tseg$ from $3$ to $39$~days in steps of $4$~days.
    \emph{(a)} Relative error $\relerr(\misFCoh,\,\misCoh)$ versus eccentricity $\ecc$.
    The dashed vertical line denotes the boundary between a ``low-$\ecc$'' range
    $\ecc\in[10^{-5},0.1]$ (left-column plots) and a ``high-$\ecc$'' range $\ecc\in[0.1,0.9]$
    (right-column plots).
    \emph{(b,c):} Relative error $\relerr$ vs observation time $\Tseg/P$.
        The solid lines in panels~(a,\,b,\,c) denote the median value, the error bars correspond to the 25th-75th percentiles,
    the circles and the stars denote the 2.5th and 97.5th percentiles, respectively.
    \emph{(d,e):} Mismatch histogram of measured mismatches $\misFCoh/\misMaxCoh$ and theoretical
    distribution in a $\Zn{6}$ lattice.
    \emph{(f,g):} Measured mismatch $\misFCoh$ versus predicted $\misCoh$.
  }
\label{fig:testing_CO_LS}
\end{figure*}

The results of these tests are shown in Fig.~\ref{fig:testing_CO_LS} for the total 20\,000 trials
performed ($2\,000$ trials for each $\Tseg$ value).
We note that the performance of the parameter-space metric starts to degrade below
$\Tseg/P \lesssim 10$ for low-eccentricity orbits [panel~$(b)$], and becomes generally poor for
high-eccentricity orbits above $\ecc\gtrsim 10^{-1}$ [panela~$(a)$ and~$(c)$].
Panels~$(d)$ and~$(e)$ show the agreement between the measured mismatch distribution
and the expected $\Zn{6}$-lattice distribution.
In panel~$(f)$ we see that the phase-metric approximation of Eq.~\eqref{eq:6a} agrees better for
small mismatch values and develops a slight tendency to \emph{over-estimates} the actual loss
$\misFCoh$ for higher mismatches.
This is a general feature seen in all four cases tested here, and qualitatively agrees with
a similar effect seen in tests of the all-sky metric for isolated CW signals
\cite{Prix:2006wm,Wette:2013wza}.

\subsubsection{Semi-coherent long-segment metric $\inc{g}^\LongSeg$}
\label{SemiCohMetTcohGreatP}

The scale for the orbital period used here is $P_0 = 2$~h, with segments of fixed length of
$\Tseg = 1$~day and varying $\Tobs$. In total we used 10 values of $\Tobs$: 1\,day, 10\,days, and
30\,days up to 100\,days in steps of 10\,days.

The parameter-space extent in orbital velocity is the same as considered in the previous
section, i.e., $d\Om \sim 1.2 \times 10^{-7}$~s$^{-1}$, resulting in a
parameters-space half-width for the period of $dP\sim 1\,$s.
Therefore $\Tseg/P \ge 12$ is satisfied for all trials.

\begin{figure*}[htbp]
  \raggedright \hspace*{0.5\columnwidth} (a)\\[-0.5cm]
  \hspace*{0.5\columnwidth}\includegraphics[width=\columnwidth]{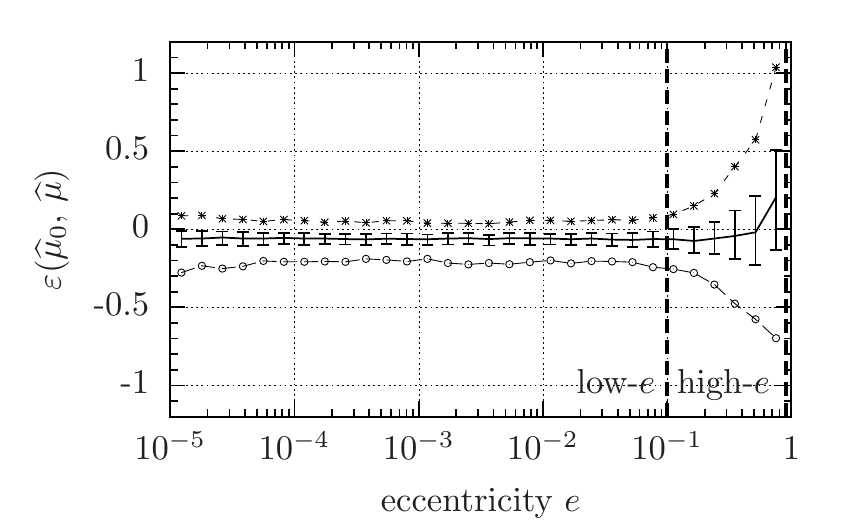}\\

  \raggedright (b)\hspace*{\columnwidth}(c)\\[-0.5cm]
  \includegraphics[clip,width=\columnwidth]{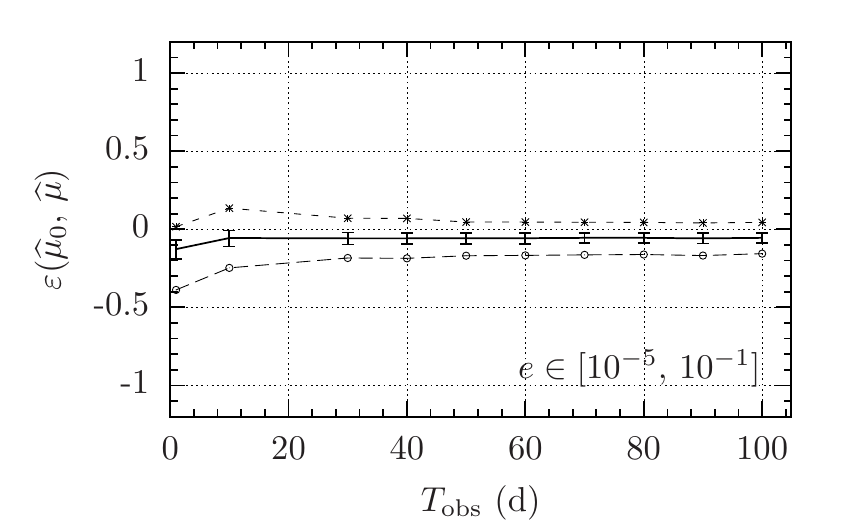}
  \includegraphics[clip,width=\columnwidth]{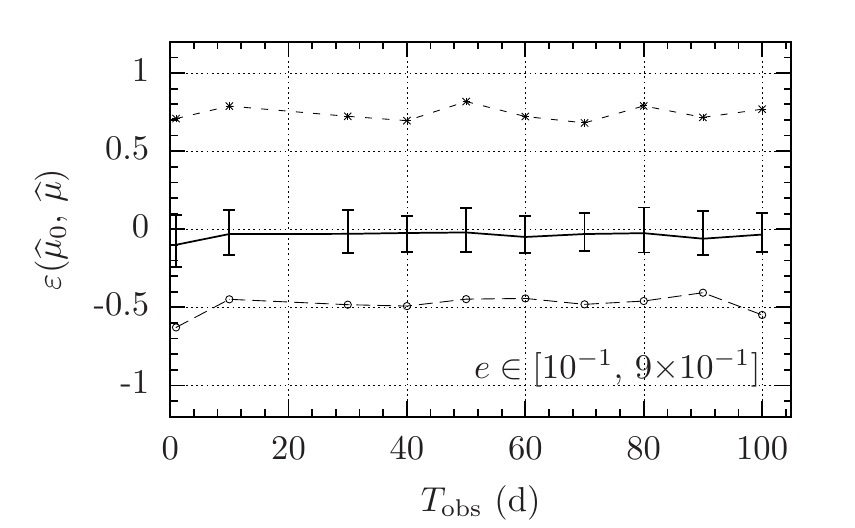}\\[-0.3cm]

  \raggedright (d)\hspace*{\columnwidth}(e)\\[-0.5cm]
  \includegraphics[clip,width=\columnwidth]{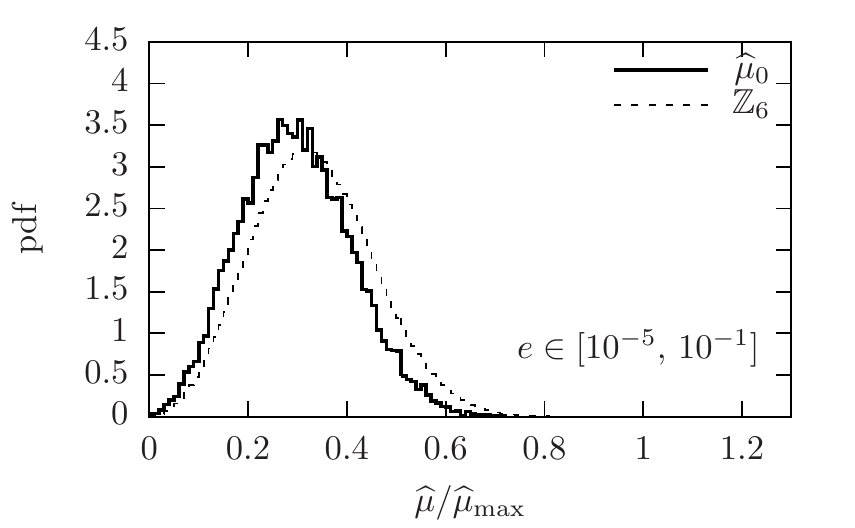}
  \includegraphics[clip,width=\columnwidth]{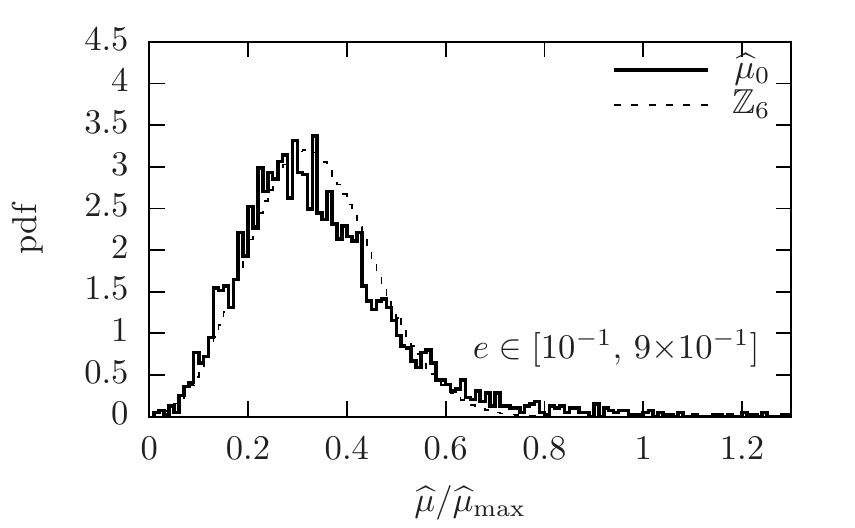}\\[-0.3cm]

  \raggedright (f)\hspace*{\columnwidth}(g)\\[-0.5cm]
  \includegraphics[clip,width=\columnwidth]{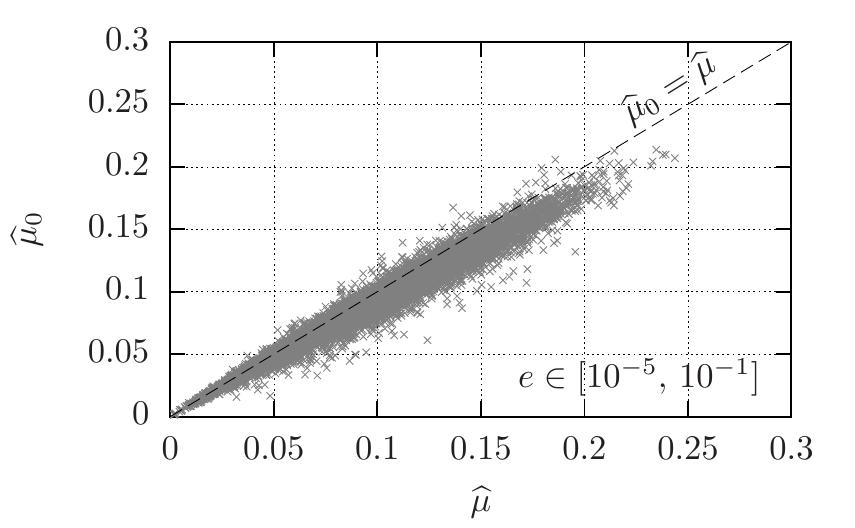}
  \includegraphics[clip,width=\columnwidth]{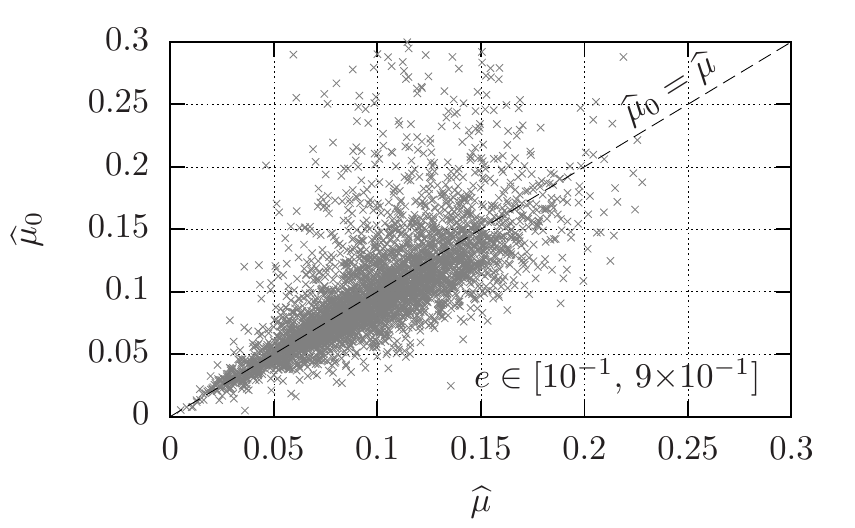}\\[-0.3cm]

  \caption{
    \emph{Semi-coherent long-segment} regime: results for metric tests on $\inc{g}^\LongSeg$ of
    Sec.~\ref{sec:semi-coherent-LS-metric} for orbital period scale of $P_0 = 2$~h and fixed segment
    length of $\Tseg = 1\,$day.
    \emph{(a)} Relative error $\relerr(\misFInc,\,\misInc)$ versus eccentricity $\ecc$.
    The dashed vertical line denotes the boundary between a ``low-$\ecc$'' range
    $\ecc\in[10^{-5},0.1]$ (left-column plots) and a ``high-$\ecc$'' range $\ecc\in[0.1,0.9]$
    (right-column plots).
    \emph{(b,c):} Relative error $\relerr$ vs observation time $\Tobs$.
            The solid lines in panels~(a,\,b,\,c) denote the median value, the error bars correspond to the 25th-75th percentiles,
    the circles and the stars denote the 2.5th and 97.5th percentiles, respectively.
    \emph{(d,e):} Mismatch histogram of measured mismatches $\misFInc/\misMaxInc$ and theoretical
    distribution in a $\Zn{6}$ lattice.
    \emph{(f,g):} Measured mismatch $\misFInc$ versus predicted $\misInc$.
  }
  \label{fig:testing_SC_LS}
\end{figure*}

The results of these tests are shown in Fig.~\ref{fig:testing_SC_LS}  for the total of 20\,000 trials
performed ($2\,000$ trials for each $\Tobs$ value).
From panels~$(a,\,b, \, c)$ we see that the agreement between measurements and predictions
is generally good in the low-eccentricity regime $\ecc<10^{-1}$, while rapidly degrading at higher eccentricites.
The case $\Tobs = \Tseg =1$~day corresponds to the single-segment coherent case, and antenna-pattern
effects are expected to play a role for observation times of order a day.

\subsection{Results in the short-segment regime}
\label{sec:results:SS}

\subsubsection{Coherent short-segment metric $\coh{g}^\ShortSeg$}
\label{CohMetSSnumSim}

There is an important technical difference in the injection algorithm in this case, as the metric is
expressed in $u$-coordinates of Eq.~\eqref{eq:ucoords} instead of physical coordinates $\Dop$.
Here the physical coordinates of an injected signal $\Dop_\sig$ are converted into $u$-coordinates,
$u_{\sig,k}\equiv u_k(\Dop_\sig)$, and then the closest lattice template $u_{\templ,k}$ is found.
In principle one could try to convert this back into physical coordinates using the expressions given in
Appendix~\ref{sec:invert-u-coord}, but this is not possible in all cases, as some templates in
$u$-coordinates do not correspond to physical coordinates and therefore cannot be inverted.
In order to circumvent this problem, we use the fact that the $u$-coordinates essentially describe
an isolated CW signal with $n-1=5$ spindown values. Hence, the perfectly-matched $\Fstat$-statistic
is computed as usual at $\Fstat_\sig\equiv\Fstat(\Dop_\sig)$, but the $\Fstat$-statistic for the mismatched template
is computed at the isolated spindown location defined by the $u$-coordinates, namely by setting
$f_\templ^{(k)} = u_{\templ,k+1}$ for $k=0,\ldots5$, i.e.\ $\Fstat_\templ\equiv\Fstat(f_\templ^{(k)})$.

The orbital period scale is set as $P_0 = 80$~days, and the resolution for the orbital angular
frequency is taken to be $d\Om = 2 \times 10^{-7}$~s$^{-1}$, and hence $P \in [62,\,97]$~days.
We sample segment lengths $\Tseg$ ranging from $2$~days up to $25$~days in steps of half a day, exploring
the range in relative segment length of $\Tseg/P\sim [0.02,\, 0.4]$, using a total of 94\,000 trials
performed (2\,000 trials for each $\Tseg$ value).

\begin{figure*}[htbp]
  \raggedright \hspace*{0.5\columnwidth}(a)\\[-0.5cm]
  \hspace*{0.5\columnwidth}\includegraphics[width=\columnwidth]{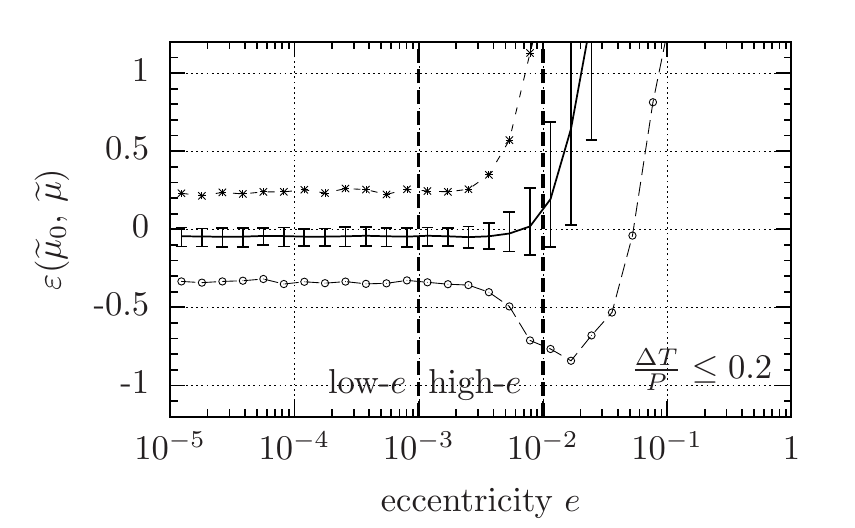}\\

  \raggedright (b)\hspace*{\columnwidth}(c)\\[-0.5cm]
  \includegraphics[clip,width=\columnwidth]{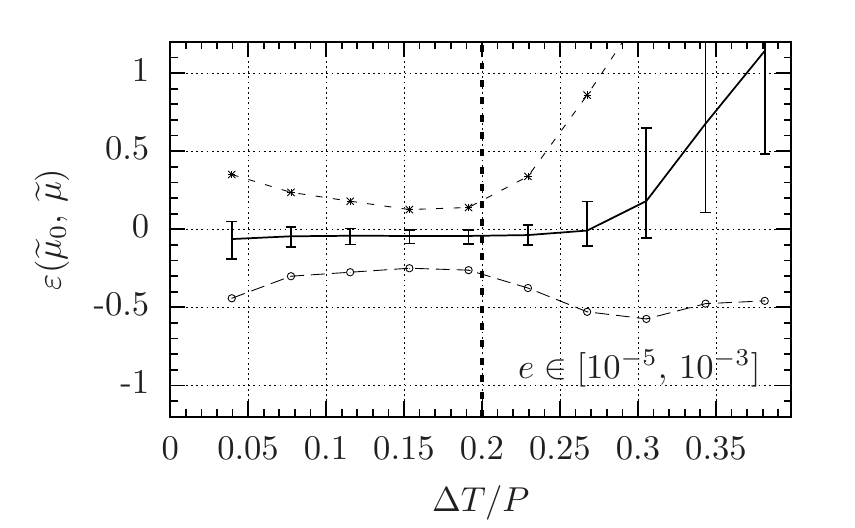}
  \includegraphics[clip,width=\columnwidth]{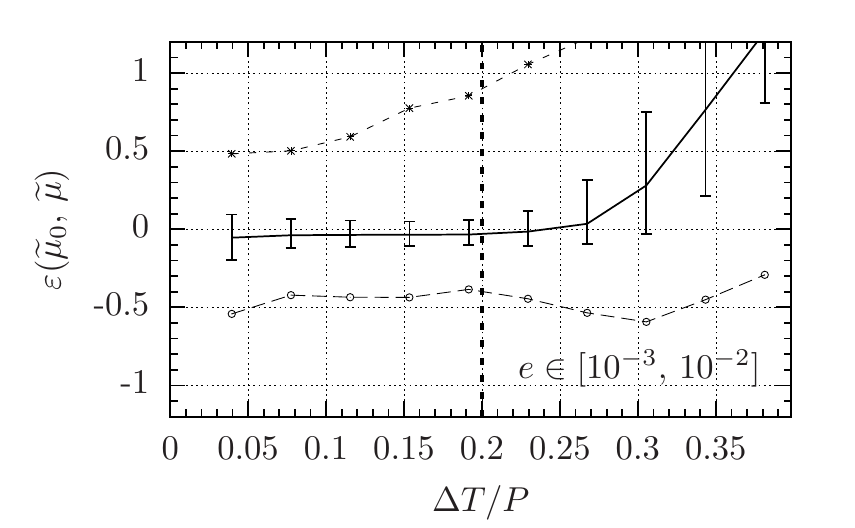}\\[-0.3cm]

  \raggedright (d)\hspace*{\columnwidth}(e)\\[-0.5cm]
  \includegraphics[clip,width=\columnwidth]{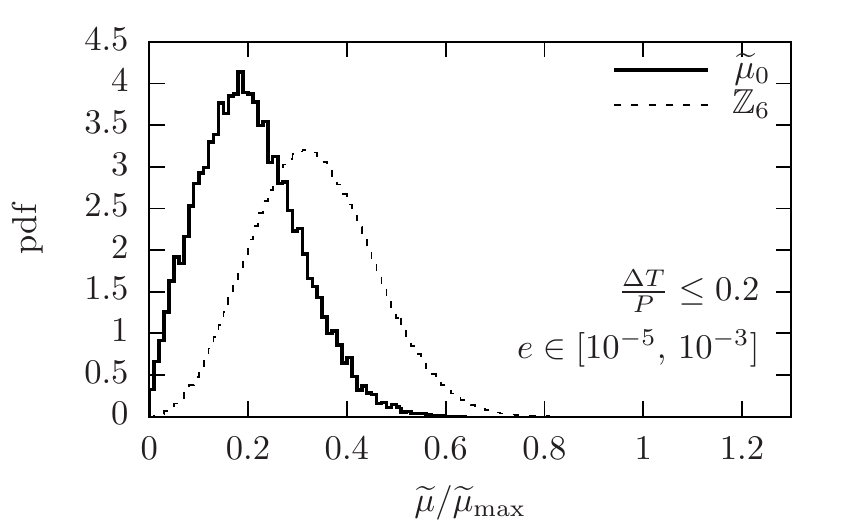}
  \includegraphics[clip,width=\columnwidth]{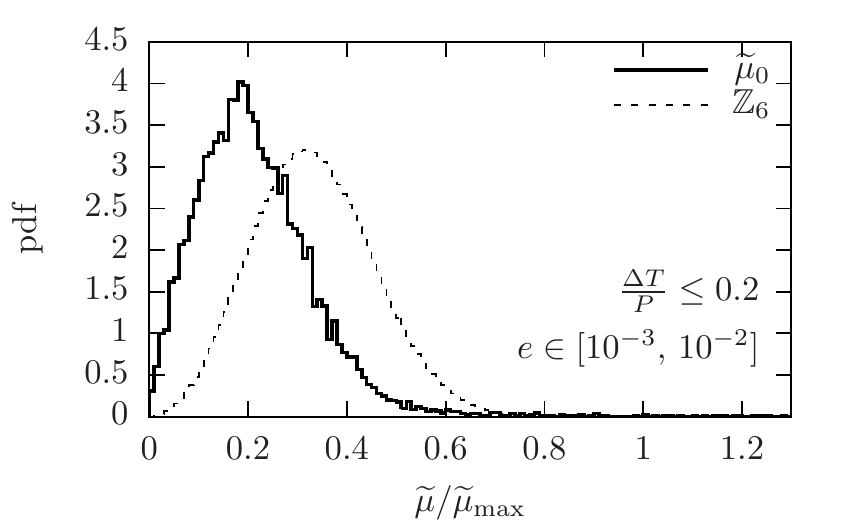}\\[-0.3cm]

  \raggedright (f)\hspace*{\columnwidth}(g)\\[-0.5cm]
  \includegraphics[clip,width=\columnwidth]{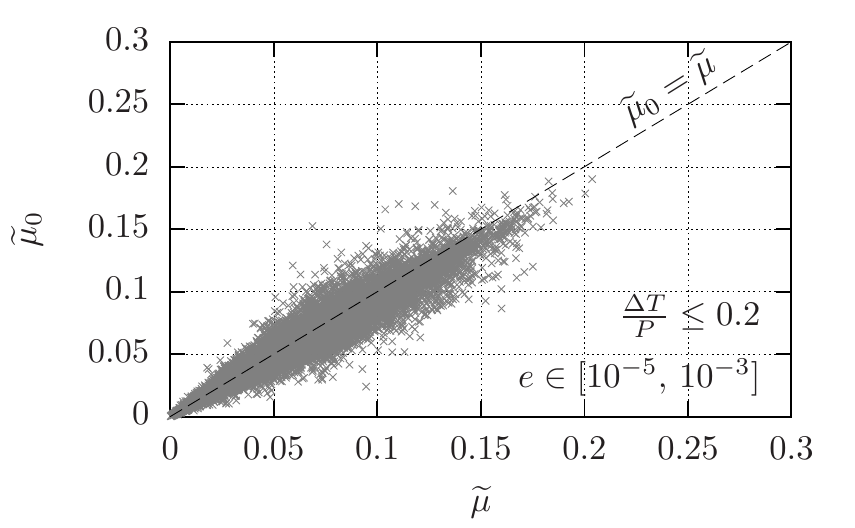}
  \includegraphics[clip,width=\columnwidth]{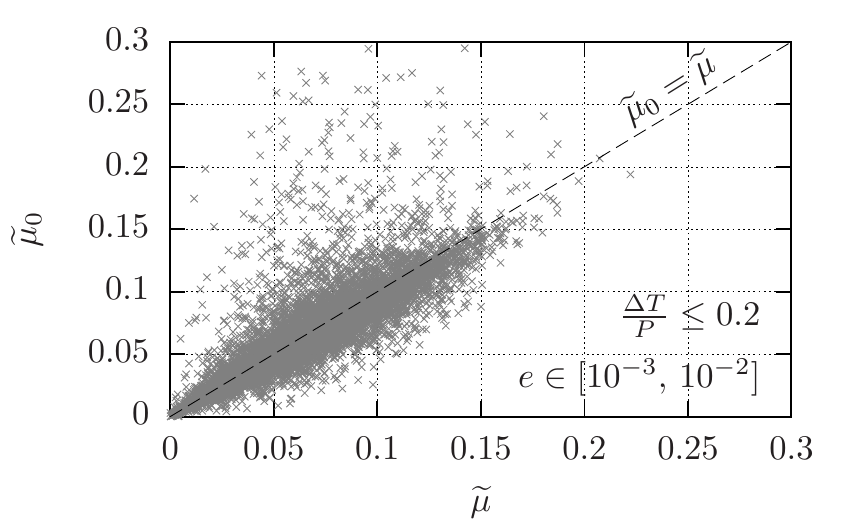}

  \caption{
    \emph{Coherent short-segment} regime: results for metric tests on $\coh{g}^\ShortSeg$ of
    Sec.~\ref{sec:coherent-SS-metric} using $u$-coordinates, for orbital period scale of $P_0 =
    80\,$days and coherent observation times in the range $\Tseg\in[2,\,25]\,$days.
    \emph{(a)} Relative error $\relerr(\misFCoh,\,\misCoh)$ versus eccentricity $\ecc$.
    The dashed vertical line denotes the boundary between a ``low-$\ecc$'' range
    $\ecc\in[10^{-5},10^{-3}]$ (left-column plots) and a ``high-$\ecc$'' range
    $\ecc\in[10^{-3},10^{-2}]$ (right-column plots).
    \emph{(b,c):} Relative error $\relerr$ vs observation time $\Tseg/P$.
            The solid lines in panels~(a,\,b,\,c) denote the median value, the error bars correspond to the 25th-75th percentiles,
    the circles and the stars denote the 2.5th and 97.5th percentiles, respectively.
    \emph{(d,e):} Mismatch histogram of measured mismatches $\misFCoh/\misMaxCoh$ and theoretical
    distribution in a $\Zn{6}$ lattice.
    \emph{(f,g):} Measured mismatch $\misFCoh$ versus predicted $\misCoh$.
    Note that in all plots except \emph{(b,c)} the range of relative segment length was restricted
    to $\Tseg/P\le0.2$, where the short-segment limit approximation is seen to be valid.
  }
  \label{fig:testing_CO_SS}
\end{figure*}

In order to better understand these results, it is interesting to consider the effective
template dimension for the different software injections, which we can quantify in terms of the highest
nonzero component $u_{\templ,k'}$ of the closest template found for an injection, namely
\begin{equation}
  \label{eq:76}
  k' \equiv \text{max } k \text{ with } |u_{\templ,k}|>0\,.
\end{equation}
\begin{figure}[htbp]
  \centering
  \includegraphics[width=\columnwidth]{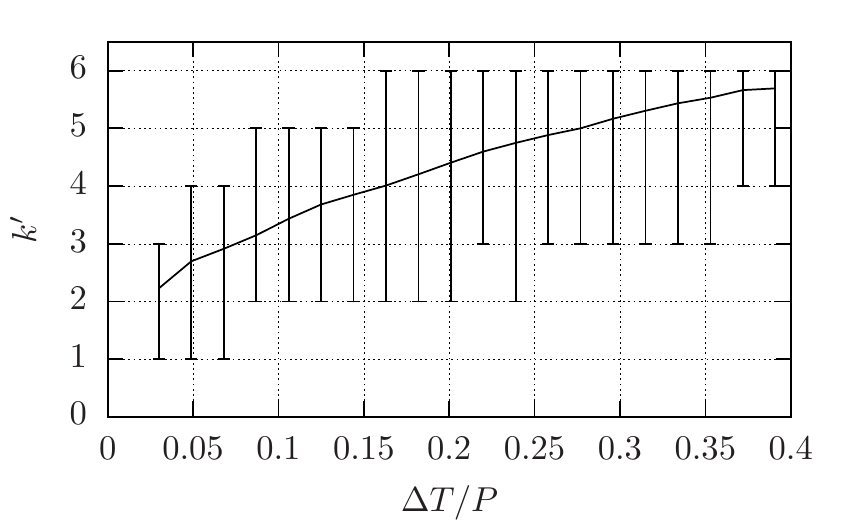}
  \caption{Effective template dimension $k'$ of Eq.~\eqref{eq:76} versus segment length $\Tseg/P$
  for the software-injection studies shown in Fig.~\ref{fig:testing_CO_SS}.
  Solid line: mean value, error bars: maximum and minimum of $k'$.}
  \label{fig:uOccupancy}
\end{figure}
This quantity is plotted Fig.~\ref{fig:uOccupancy} for the injections performed for
Fig.~\ref{fig:testing_CO_SS}, giving the range of effective template dimensions $k'$ as a function
of relative segment length $\Tseg/P$.
As expected, we see that the effective template dimension increases with $\Tseg/P$ and eventually would require
more than 6 $u$-coordinates for $\Tseg/P\gtrsim 0.2$, corresponding to the region where the
$u$-metric performance is seen in Fig.~\ref{fig:testing_CO_SS}(b) to start to degrade.
We also see that the average template-bank dimension is not constant and generally
less than 6, which explains the discrepancy between the sampled mismatch distribution in
Fig.~\ref{fig:testing_CO_SS}(d) compared to the expected $\Zn{6}$-lattice distribution.

The results of the injection tests are shown in Fig.~\ref{fig:testing_CO_SS}.
From panels~$(b)$ and~$(c)$ we see that the approximation in terms of
6 $\uCoord$-coordinates used here starts to break down in the range
$\Tseg/P\gtrsim[0.25,0.35]$, as anticipated from Eq.~\eqref{eq:63} and Fig.~\ref{fig:uOccupancy}.
Also, contrary to the other three limits tested, the small-eccentricity approximation only seems to
work well for eccentricities up to about  $\ecc\lesssim 3\times10^{-3}$, after which it rapidly
deteriorates and increasingly loses predictive power beyond $\ecc\gtrsim10^{-2}$ [see panel~$(a)$].

\subsubsection{Semi-coherent short-segment metric $\inc{g}^\ShortSeg$}
\label{SemicohMetSSnumSim}

\begin{figure*}[htbp]

  \raggedright \hspace*{0.5\columnwidth}(a)\\[-0.5cm]
  \hspace*{0.5\columnwidth}\includegraphics[width=\columnwidth]{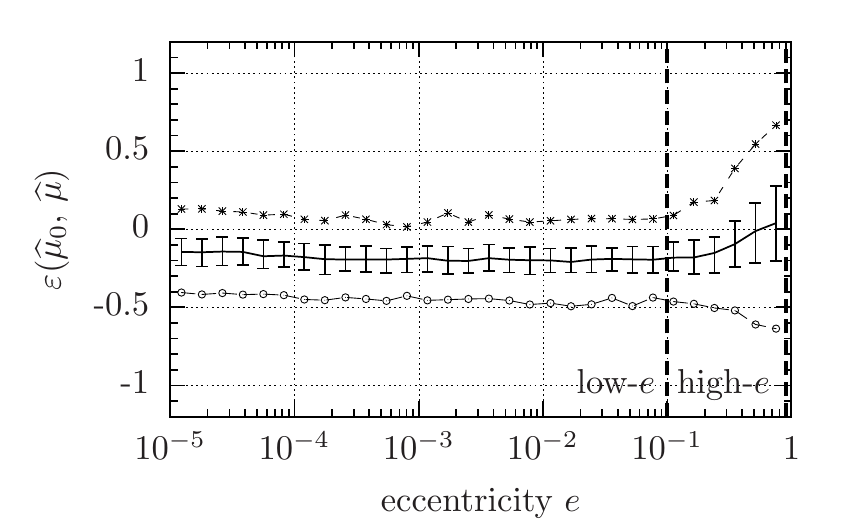}\\

  \raggedright (b)\hspace*{\columnwidth}(c)\\[-0.5cm]
  \includegraphics[clip,width=\columnwidth]{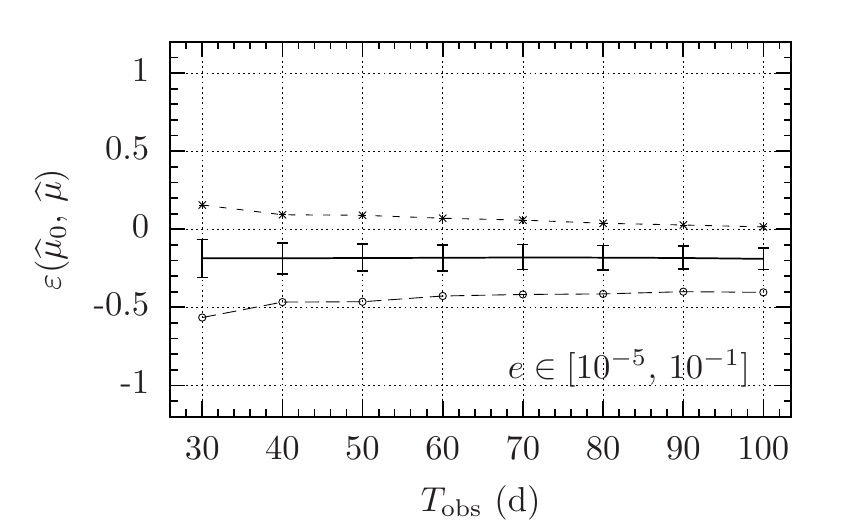}
  \includegraphics[clip,width=\columnwidth]{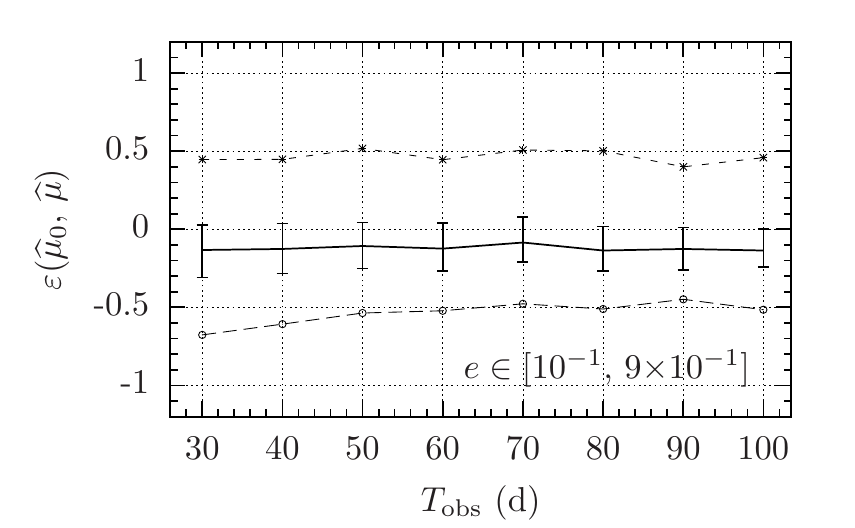}\\[-0.3cm]

  \raggedright (d)\hspace*{\columnwidth}(e)\\[-0.5cm]
  \includegraphics[clip,width=\columnwidth]{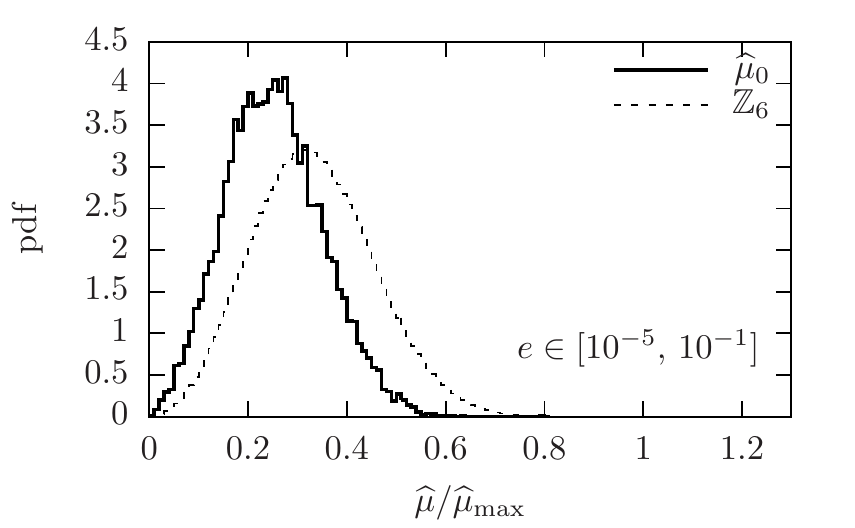}
  \includegraphics[clip,width=\columnwidth]{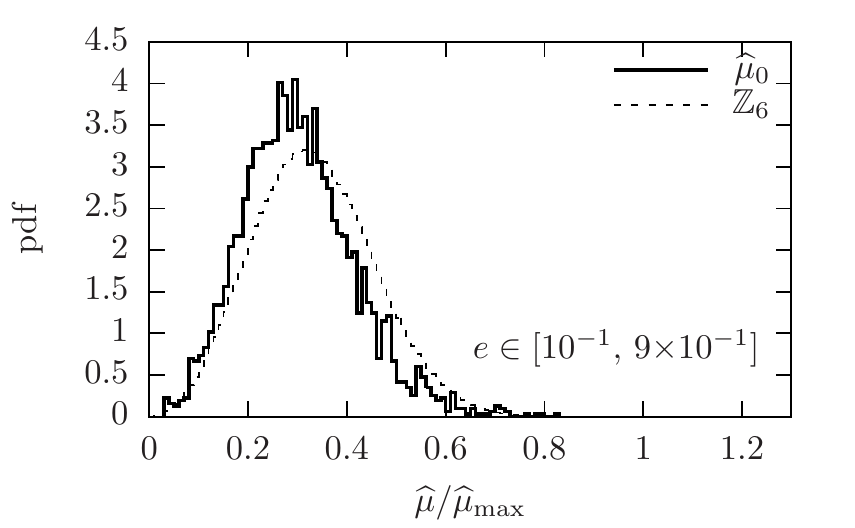}\\[-0.3cm]

  \raggedright (f)\hspace*{\columnwidth}(g)\\[-0.5cm]
  \includegraphics[clip,width=\columnwidth]{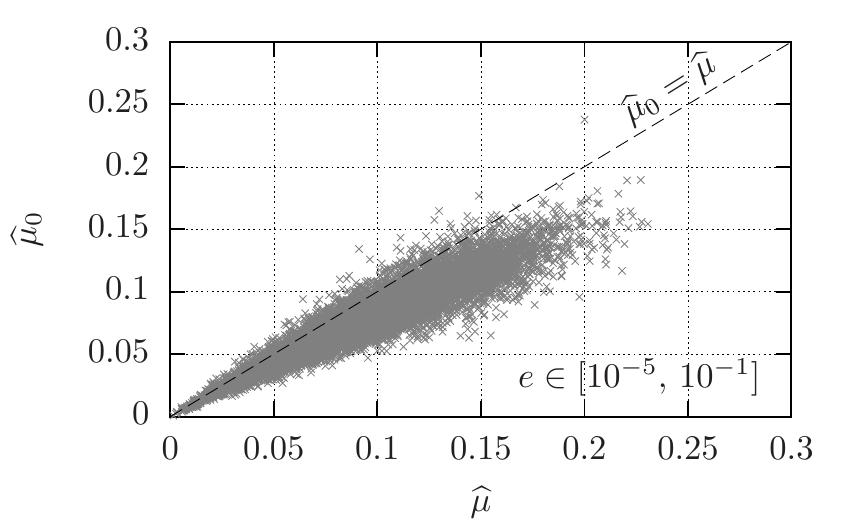}
  \includegraphics[clip,width=\columnwidth]{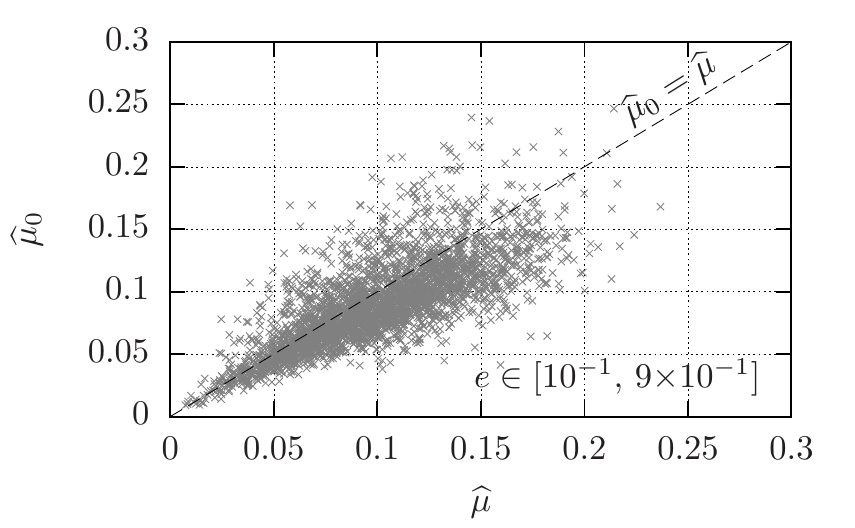}

  \caption{
    \emph{Semi-coherent short-segment} regime: results for metric tests on $\inc{g}^\ShortSeg$ of
    Sec.~\ref{sec:semi-coherent-SS-metric} for orbital period scale of $P_0 = 10$~days, and a fixed
    segment length of $\Tseg = 1$~day.
    \emph{(a)} Relative error $\relerr(\misFInc,\,\misInc)$ versus eccentricity $\ecc$.
    The dashed vertical line denotes the boundary between a ``low-$\ecc$'' range
    $\ecc\in[10^{-5},0.1]$ (left-column plots) and a ``high-$\ecc$'' range $\ecc\in[0.1,0.9]$
    (right-column plots).
    \emph{(b,c):} Relative error $\relerr$ vs observation time $\Tobs$.
            The solid lines in panels~(a,\,b,\,c) denote the median value, the error bars correspond to the 25th-75th percentiles,
    the circles and the stars denote the 2.5th and 97.5th percentiles, respectively.
    \emph{(d,e):} Mismatch histogram of measured mismatches $\misFInc/\misMaxInc$ and theoretical
    distribution in a $\Zn{6}$ lattice.
    \emph{(f,g):} Measured mismatch $\misFInc$ versus predicted $\misInc$.
  }
\label{fig:testing_SC_SS}
\end{figure*}

The scale for the orbital period here is fixed to $P_0 = 10$~days, and we used a fixed segment
length of $\Tseg = 1$~day.
The parameter space extent in terms of $\Om$ is
$d\Om=2\,\max[( \inc{g}^{\ShortSeg}_{\Om\Om})^{-1/2}]\sim6.6\times 10^{-8}$~s$^{-1}$
resulting in $d P\sim 2.2\,$hours and therefore we have at most $\Tseg/P\lesssim 0.12$.
We use a varying $\Tobs$ ranging from 30 up to 100 days in steps of 10 days.
The results of these tests are shown in Fig.~\ref{fig:testing_SC_SS} for the total 16\,000 trials
performed ($2\,000$ trials for each $\Tobs$ value).
We note that the validity extent of the semi-coherent metric is constantly acceptable
for low-eccentricity orbits [panel~$(b)$] and degrades (albeit much less than in the other three
cases) for higher eccentricity $\ecc\gtrsim 10^{-1}$ [panels~$(a)$ and~$(c)$].
Panels~$(d)$ and~$(e)$ compare the mismatch histogram of measured mismatches with the
theoretical distribution in a $\Zn{6}$ lattice.
Furthermore, we also note that in this regime the metric mismatch $\misInc$ tends to generally
\emph{over-estimate} $\misFInc$ somewhat more strongly than in the other cases [see panels~$(f)$ and~$(g)$].

\subsection{General discussion of metric testing results}
\label{sec:gener-comm-monte}

Summarizing the results presented in Figs.~\ref{fig:testing_CO_LS}-~\ref{fig:testing_SC_SS}, we
observe that the agreement between metric mismatch predictions and the measured relative
loss in $\Fstat$-statistic is generally very good (typically no worse than
$\sim 10-30\%$) within its range of applicability. In the following we further quantify where various
approximations start to break down.

The first-order small-eccentricity approximation (see Sec.~\ref{sec:small-eccentr-appr})
starts to noticeably degrade only above $\ecc \gtrsim 0.1$ [Fig.~\ref{fig:testing_CO_LS}(a),
Fig.~\ref{fig:testing_SC_LS}(a) and Fig.~\ref{fig:testing_SC_SS}(a)] except in the coherent
short-segment regime where it breaks down above $\ecc \gtrsim 5\times10^{-3}$ [Fig.~\ref{fig:testing_CO_SS}(a)].
Further investigation of this behaviour might be interesting, but is beyond the scope of this study.

The (phase-) metric approximation Eq.~\eqref{eq:6a} is expected from previous studies
\cite{Prix:2006wm,Wette:2013wza} to be quite accurate (in suitable coordinates) up to values of
mismatch of $\mis\lesssim0.2$, after which higher-order terms start to become more noticeable,
which tend to \emph{reduce} the actual measured mismatches compared to the metric predictions.
While the nonlinear regime is hardly explored here, this general trend can still be seen to
some extent in panels (f) of
Figs.~\ref{fig:testing_CO_LS},~\ref{fig:testing_SC_LS},~\ref{fig:testing_CO_SS},~\ref{fig:testing_SC_SS}.

\section{Scorpius~X-1 sensitivity estimate}
\label{sec:sco-x1-sensitivity}

We can use the metric expressions and template counts derived in Sec.~\ref{sec:number-templates} to estimate
the optimal achievable sensitivity of a semi-coherent search directed at Scorpius~X-1.

\subsection{Torque-balance level}
\label{sec:torque-balance-level}

In order to quantify the sensitivity of a search \emph{independent} of the detector noise floor, we
can define the \emph{sensitivity depth} \cite{2014arXiv1410.5997B} as
\begin{equation}
  \label{eq:79}
  \SensDepth{C}{\pFA} \equiv \frac{\sqrt{\Sn}}{h^C_{\pFA}}\,,
\end{equation}
in terms of the harmonic mean over detector noise floors $\Sn$, and the strain sensitivity (or upper
limit) $h^C_{\pFA}$ of a search at a certain confidence level $C$ (i.e.\ detection probability) and
false-alarm threshold $\pFA$. This quantity (with dimensions of $\Hz^{-1/2}$) is useful as a simple
and intuitive measure for how far below the noise floor a given search setup can reach.

An interesting astrophysical model postulating torque balance between the accretion and CW emission
\cite{papaloizou78:_gravit,wagoner84:_gravit,bildsten98:_gravit} yields a predicted CW amplitude
(assuming a NS with 10\,km radius and a mass of $1.4\,M_{\odot}$) for Scorpius~X-1 of
\begin{equation}
  \label{eq:71}
  \hO \sim 3.5\times10^{-26} \sqrt{\frac{300\,\Hz}{\nu}}\,,
\end{equation}
where $\nu$ is the (unknown) NS spin frequency.

\begin{figure}[htbp]
  \centering
  \includegraphics[width=\columnwidth,clip]{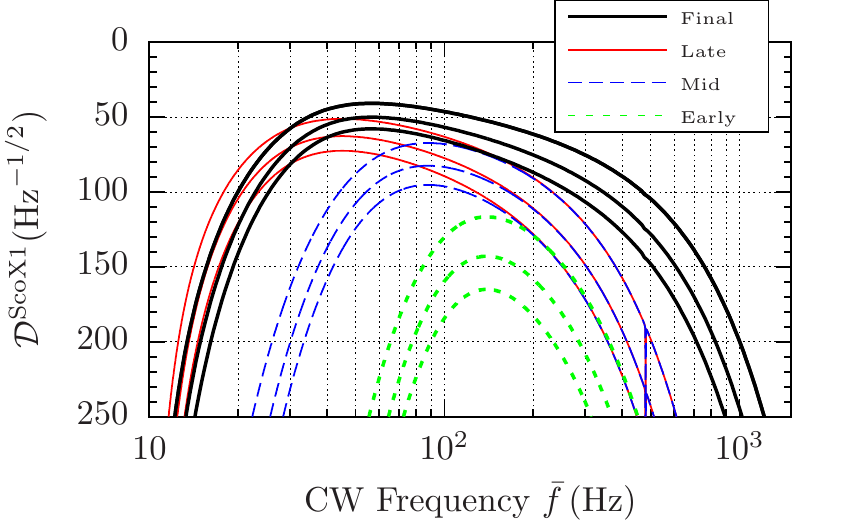}
  \caption{Sensitivity depth $\SensDepth{\ScoX}{}$ of Eq.~\eqref{eq:79} for \mbox{Scorpius~X-1} torque-balance
    level Eq.~\eqref{eq:71} versus CW frequency $\fbar$.
    Different line styles correspond to different progressive aLIGO sensitivities
    (``Early'', ``Mid'', ``Late'' and ``Final'')~\cite{aLIGO_scenarios,2013arXiv1304.0670L,LSC:_aLIGO_sensitivity}.
    For each detector sensitivity three different emission models are shown, namely
    \emph{top:} ``mountain'' ($\fbar=2\nu$),
    \emph{mid:} r-mode ($\fbar\approx4\nu/3$),
    \emph{bottom:} precession ($\fbar\approx\nu$), respectively.
    }
  \label{fig:minSearchDepth}
\end{figure}

Figure~\ref{fig:minSearchDepth} shows the minimum required sensitivity depth $\SensDepth{\ScoX}{}$
to reach the Scorpius~X-1 torque-balance limit of Eq.~\eqref{eq:71} for different emission models
(``mountain'' deformation with $\fbar = 2\,\nu$, r-mode emission with $\fbar\approx 4\nu/3$, and precession with
$\fbar\approx\nu$), assuming different aLIGO sensitivity curves  (``Early'', ``Mid'', ``Late'' and ``Final'')
\cite{LSC:_aLIGO_sensitivity,aLIGO_scenarios,2013arXiv1304.0670L}.

For all the following sensitivity estimates, we will assume a detector duty cycle of $80\%$
(following \cite{2013arXiv1304.0670L}), a  $90\%$ confidence level (i.e.\ detection probability), and
a false-alarm level of $\pFA=10^{-10}$ (appropriate for a first-stage wide-parameter search).

\subsection{Best theoretically achievable sensitivity}
\label{sec:upper-bounds-theor}

\subsubsection*{Short-segment regime}
\label{sec:short-segment-regime}

Let us first consider the short-segment regime, assuming a ``best-case'' search with segments of length
$\Tseg = 4.7\,\Hours$ (i.e. roughly 1/4 of the Scorpius~X-1 orbital period, quoted in Table~\ref{tab:Scorpius-X1MDCvalues}),
total timespan $\Tobs=365\,\Days$ (therefore $\Nseg=1864$
segments), 2 detectors [LIGO H and Livingston (L)], and a very fine search grid of
average mismatch $\mis=0.01$. Using the method outlined in \cite{Wette:2011eu} and implemented in
\cite{OctApps}, we obtain a resulting sensitivity depth of
$\SensDepth{90\%}{\mathrm{1e-10}}\sim 56\,\Hz^{-1/2}$.
Assuming 3 equal-sensitivity detectors [LIGO H, L and Virgo (V)], this value increases to
$\SensDepth{90\%}{\mathrm{1e-10}}\sim 68\,\Hz^{-1/2}$.

\subsubsection*{Long-segment regime}
\label{sec:long-segment-regime}

In the long-segment regime of Sec.~\ref{sec:long-segment-limit}, the maximal segment length
would be restricted by the astrophysical concern of ``spin wandering'', namely a stochastic
variability of the spin frequency due to variations in the accretion rate.
There is substantial astrophysical uncertainty
\cite{Brady:1998nj,2000AIPC..523...65U,2014arXiv1412.0605T} about the details of this process and
its magnitude, which is beyond the scope of this work.
We consider assumptions roughly similar to those given in the recent Scorpius~X-1 MDC
\cite{ScoX1:MDC1}, which posited a random frequency derivative of order
$|\fdot| \le \fdot_s \sim 10^{-12}\,\Hz/\Sec$, changing on a timescale of
$t_s\sim10^6\,\Sec$. Over this timescale, the frequency drift would therefore be
$\delta \coh{f}_s\le \fdot_s\,t_s\sim 10^{-6}\,\Hz$. Over a total observation time $\Tobs$, this can be
modelled as a random walk with expected total drift (with respect to the midpoint in time) of order
$\delta\inc{f}_s \sim \sqrt{\Tobs/(2t_s)}\,\delta\coh{f}_s$.
The maximal coherent segment length could therefore
be estimated roughly by the requirement that the total frequency drift should be less than the frequency
resolution $\sim 1/\Tseg$ of the search, in order to avoid any significant loss of SNR.
This yields the constraint
\begin{equation}
  \label{eq:81}
  \Tseg \lesssim \sqrt{2}\,\left(\sqrt{\Tobs\,t_s}\,\fdot_s\right)^{-1}\,.
\end{equation}
Assuming $\Tobs=1\,$year and the above MDC spin-wandering model, we find $\Tseg\lesssim3\,\Days$.
Given the substantial uncertainty in these parameters, for comparison we also consider a more
optimistic scenario of $\Tseg\lesssim10\,\Days$.

The best achievable sensitivity for $\Tseg=3\,\Days$ and $\Nseg=120$ segments, assuming a
small mismatch of $\mis=0.01$, can be estimated for a 2-detector network as
$\SensDepth{90\%}{\mathrm{1e-10}}\sim 105\,\Hz^{-1/2}$, and for 3 detectors we find
$\SensDepth{90\%}{\mathrm{1e-10}}\sim 127\,\Hz^{-1/2}$.

The best sensitivity that can be achieved for $\Tseg=10\,\Days$ and $\Nseg=36$ segments, and
$\mis=0.01$, can be estimated for a 2-detector network as
$\SensDepth{90\%}{\mathrm{1e-10}}\sim 134\,\Hz^{-1/2}$,
and for 3 detectors we find
$\SensDepth{90\%}{\mathrm{1e-10}}\sim 163\,\Hz^{-1/2}$.

In the following we will focus only on the long-segment regime, and assess the required
computing power as well as the resulting sensitivity depth that can actually be reached in practice.

\subsection{Scorpius~X-1 search parameter space}
\label{sec:number-templates-1}

We assume the Scorpius~X-1 parameters and uncertainties given in Table \ref{tab:Scorpius-X1MDCvalues}.
Note that, contrary to the Scorpius~X-1 MDC~\cite{ScoX1:MDC1} and previous searches
\cite{2014arXiv1412.0605T,2014arXiv1412.0605T}, we also allow for a nonzero uncertainty on the
eccentricity, in addition to the circular-orbit assumption that is also not ruled out by
current observations~\cite{ScoX1eccValue}.

\begin{figure}[htbp]
  \centering
  \includegraphics[width=\columnwidth,clip]{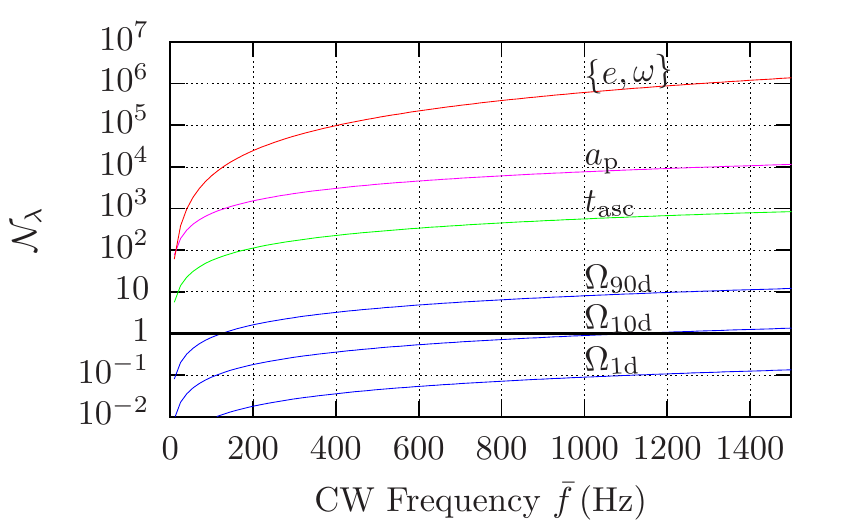}
  \caption{Effective number of (coherent or semi-coherent) templates per dimension $\NtPerDim{\Dop}$
    of Eq.~\eqref{eq:90} (in the long-segment regime) as function of CW frequency $\fbar$ for the Scorpius~X-1 parameter space given in
    Table~\ref{tab:Scorpius-X1MDCvalues}, using $\pm 3$ sigma ranges.
    The template numbers are evaluated for $\misMax=0.1$ and mean values of $\asini$ and $\Om$.
    Note that assuming a $2\pi$ uncertainty in $\argp$ entails
    $\NtPerDim{\{\ecc,\argp\}} = \NtPerDim{\ecc}^2\,\pi$.
    The template count $\NtPerDim{\Om}$ is the only one depending on the observation time $T$ and is
    shown for $T = 1\,\Days,\,10\,\Days,\,90\,\Days$.
  }
  \label{fig:templates_per_dim_ScoX1}
\end{figure}

\begin{table}
\caption{\label{tab:Scorpius-X1MDCvalues}  Assumed Scorpius~X-1 system parameters and uncertainties.
  Ranges of the form $\Dop \pm \Delta\Dop$ denote the mean and Gaussian 1-sigma
  uncertainty, while $[\Dop_0,\Dop_1]$ denotes a uniform probability range.
  The time of ascending node $\tAsc$ was computed from the originally-measured time of inferior
  conjunction $T_0$ \cite{2014ApJ...781...14G} via $\tAsc = T_0 - P/4$. The projected semi-major
  axis $\asini$ is related to the projected radial maximal velocity $K_1$, appearing in~\cite{2002ApJ...568..273S},
  by $\asini = K_1\,P/(2\pi c)$.
}
\begin{ruledtabular}
\begin{tabular}{l l r}
Parameter & Value & Ref.\\
\hline
$\asini\,(\Sec)$                & $1.44 \pm 0.18$         & \cite{2002ApJ...568..273S,2014arXiv1412.0605T}\\
$\tAsc\,(\mathrm{GPS}\,\Sec)$ & $897753994 \pm 100$     & \cite{2014ApJ...781...14G}\\
$P\,(\Sec)$                      & $68023.70496 \pm 0.0432$    & \cite{2014ApJ...781...14G}\\
$\ecc$                            & $(\approx 0) \text{ or } (0.033 \pm 0.018)$ & \cite{ScoX1eccValue}\\
$\argp\,$(rad)                   & $[0,\,2\pi]$ & \cite{ScoX1eccValue}\\
\end{tabular}
\end{ruledtabular}
\end{table}
Assuming 3-sigma ranges for the Scorpius~X-1 parameter uncertainties, the resulting per-dimension template
numbers $\NtPerDim{\Dop^i}$ at maximal mismatch $\misMax=0.1$ of Eq.~\eqref{eq:90} are shown in
Fig.~\ref{fig:templates_per_dim_ScoX1}.
We see that, for the potential uncertainty on eccentricity $\ecc$ of Table
\ref{tab:Scorpius-X1MDCvalues}, one would have to include $\ecc$ and $\argp$ as effective search
dimensions in the long-segment limit.
On the other hand, the assumption of full circularization $\ecc\approx0$ is not ruled out by the
current observations and therefore also constitutes a reasonable alternative model to investigate.

The only dimension where the resolution depends on the timespan $T$ is
$\Om$. We see, for example, that for coherent times $\Tseg\lesssim 10\,\Days$ the uncertainty in $\Om$ will
typically not be resolved by the coherent metric (as long as $\misMaxCoh\ge0.1$), as
$\NtPerDim{\Om}(T<10\Days)\lesssim 1$. In this case one could simply exclude this search dimension
in the coherent template count $\coh{\Nt}$ of Eq.~\eqref{eq:NumTemplates}.
On the other hand, the semi-coherent fine grid (assuming $\Tobs \ge 90\,\Days$ and
$\misMaxInc\le0.1$) can typically resolve this dimension over some frequency range where $\NtPerDim{\Om}(T>90\Days)>1$.
In both cases, however, this depends on the resulting mismatch parameter $\misMax$, which is itself
an optimization parameter. Therefore, as seen in
Fig.~\ref{fig:templates_per_dim_ScoX1}, one potentially has to deal with situations where the
$\Om$-dimension will be unresolved at lower frequencies but resolved at higher frequencies.
One way to obtain a correct template estimate in such a case would be to determine if a transition
frequency exists, where the $\Om$-dimension changes from unresolved to resolved, then stitch together
the respective correct template counts over each frequency range. Alternatively, one can simply split
the frequency range into narrow slices, such that the dimensionality can be assumed constant over
each slice, and sum up the respective number of templates.

Note that due to the large separation $\Dma$ between the orbital epoch $\tAsc$ in
Table~\ref{tab:Scorpius-X1MDCvalues} and the aLIGO gravitational wave search epoch, we have to consider the
potential increase in uncertainty on $\tAsc$, as discussed in \cite{2014ApJ...781...14G}.
As discussed in Sec.~\ref{sec:num-templates-long-segm}, this is not a concern for a grid where both
$\Om$ and $\tAsc$ are fully resolved by the metric, as is expected to be the case of the
semi-coherent grid. On the other hand, the coherent grids will typically not resolve the $\Om$ dimension, and
therefore an increase of about 2.5--3 in uncertainty on $\tAsc$ would have to be taken into account.
However, for the present study, we assume (as seems likely \cite{2014ApJ...781...14G}) that there will be further
observations on Scorpius~X-1 closer to the aLIGO epoch, which will again constrain the uncertainty on $\tAsc$ to a
level similar (or better) than that given in Table~\ref{tab:Scorpius-X1MDCvalues}.

\subsection{Computing cost model}
\label{sec:computing-cost-model}

As described in \cite{Prix:2012yu}, a semi-coherent StackSlide search (introduced in
Sec.~\ref{sec:semi-coher-detect}) for a fixed parameter space has the following tuneable
parameters: the segment length $\Tseg$, the number of segments $\Nseg$, the maximal
template-bank mismatch for the (semi-coherent) fine grid $\misMaxInc$, and for the per-segment
(coherent) coarse grids $\misMaxCoh$.
The aim is therefore to maximize the resulting sensitivity over this space of search parameters, under the
constraint of fixed total computing cost $\Ctot(\misMaxCoh,\misMaxInc,\Nseg,\Tseg) = \Cost_0$.

The total computing cost $\Ctot$ of the StackSlide $\Finc$-statistic can be written as~\cite{Prix:2012yu}
\begin{equation}
\label{eq:Ctot}
\Ctot(\misMaxCoh,\misMaxInc,\Nseg,\Tseg)  = \CCoh + \CInc\,,
\end{equation}
where $\CCoh(\misMaxCoh,\Nseg,\Tseg)$ is the cost of $\Nseg$ per-segment coherent
$\Fcoh$-statistic searches over a coarse grid with maximal mismatch $\misMaxCoh$,
and $\CInc(\misMaxInc,\Nseg,\Tseg)$ is the cost of incoherently summing (and interpolating) these
$\Fcoh_\iSeg$-values across all segments on a fine grid with maximal mismatch $\misMaxInc$.

The coherent cost for the $\Fcoh$-statistic computation can be expressed as
\begin{equation}
  \label{eq:Ccoarse}
  \CCoh = \Nseg\,\NtCoh(\misMaxCoh,\Tseg)\,\Ndet\;\coh{c}_1(\Tseg) \,,
\end{equation}
where  $\Ndet$ is the number of detectors, and $\coh{c}_1$ is the (per-detector) computing cost per
template, which depends on the algorithm used to compute the $\Fcoh$-statistic.
For the (generally slower) SFT-based demodulation method~\cite{prix:_cfsv2},
this can be expressed (for gapless data) as
\begin{equation}
  \label{eq:72}
  \coh{c}_1^\SFT(\Tseg) = \coh{c}_0^\SFT\,\frac{\Tseg}{\Tsft}\,,
\end{equation}
where $\coh{c}_0^\SFT \approx \,4\times\,10^{-8}$~s is an implementation- and hardware-dependent
fundamental computing cost for binary-CW templates (measured by timing the current $\Fcoh$-statistic
LALSuite code~\cite{LALSuite}).

Using instead the more efficient Fast Fourier Transform (FFT) based ``resampling'' method
\cite{Jaranowski:1998qm,2010PhRvD..81h4032P}, we find the time per template $\coh{c}_1$ to be
approximately constant (for search frequency bands of $\gtrsim 10^5$ frequency bins) as
\begin{equation}
  \label{eq:74}
  \coh{c}_1^\Resamp \approx 3\times10^{-7}\,\Sec\,.
\end{equation}
The semi-coherent cost of summing across segments can be expressed as
\begin{equation}
  \label{eq:Cfine}
  \CInc  = \Nseg\,\NtInc(\misMaxInc,\Tseg,\Nseg)\,\inc{c}_0,
\end{equation}
where $\inc{c}_0 \approx5\times\,10^{-9}$~s is an implementation- and hardware-dependent fundamental
computing cost of adding one value of $2\,\Fcoh_{\iSeg}$ for one fine-grid point (measured by timing
the LALSuite \texttt{HierarchSearchGCT} code~\cite{LALSuite}).

Note that the sensitivity estimates presented in
Tables~\ref{tab:OptimalSetupsEaH-Demod}--~\ref{tab:OptimalSetupsEaH-ResampEcc_v2} are using
``Einstein@Home months'' ($\mathrm{EM}$) as a computing-cost unit. This unit is defined as
$12\,000$\,core-months\footnote{Counting only the 30\%-50\% CPUs currently devoted to gravitational wave searches,
  and including the double-computation of all results for validation.} on a CPU that achieves the
above-quoted timings $\inc{c}_0$ and $\coh{c}_1$.
Empirically this corresponds to the performance of the ``average CPU'' currently participating in
Einstein@Home, and is comparable to roughly an Intel Core i7-2620M or Intel Xeon E3-1220 v3.

\subsection{Sensitivity estimates}
\label{sec:sco-x-1}

With the computing-cost model at hand we can employ the semi-analytic machinery of \cite{Prix:2012yu} to
find optimal StackSlide setups for given computing cost.
We can then estimate the corresponding sensitivity of each setup with the fast and accurate method
developed in \cite{Wette:2011eu}. Both the optimization and sensitivity-estimation algorithms have
been implemented in \cite{OctApps} and are used here.

Due to the rapidly increasing computing cost with frequency [see Eqs.~\eqref{eq:31N}--\eqref{eq:86}], it
is important to limit the frequency range searched in some way.
Hence, we focus on constructing example setups that beat the torque-balance level (shown in
Fig.~\ref{fig:minSearchDepth}) over their respective frequency range.

We consider the two different constraints on the segment length $\Tseg$ due to spin wandering, as
discussed in Sec.~\ref{sec:long-segment-regime}, namely $\Tseg\le10\,\Days$ and $\Tseg\le3\,\Days$,
respectively.
For $\Tseg\le10\Days$, the results for different $\Fstat$-computation methods and assumptions about
eccentricity are summarized in Table~\ref{tab:OptimalSetupsEaH-Demod} (for SFT-based demodulation
and zero uncertainty on $\ecc$), Table~\ref{tab:OptimalSetupsEaH-ResampCirc} (for FFT-based
resampling and zero uncertainty on $\ecc$), and Table~\ref{tab:OptimalSetupsEaH-ResampEcc}
(for FFT-based resampling and 3-sigma uncertainty of $\Delta\ecc=0.087$).

Similarly, for $\Tseg\le3\Days$, the results for different $\Fstat$-computation methods and
assumptions about eccentricity are summarized in Table~\ref{tab:OptimalSetupsEaH-Demod_v2} (for SFT-based demodulation
and zero uncertainty on $\ecc$), Table~\ref{tab:OptimalSetupsEaH-ResampCirc_v2} (for FFT-based
resampling and zero uncertainty on $\ecc$), and Table~\ref{tab:OptimalSetupsEaH-ResampEcc_v2}
(for FFT-based resampling and 3-sigma uncertainty of $\Delta\ecc=0.087$).

In each table the first line shows a setup beating the $\fbar=2\nu$ torque-balance depth for the ``Final''
aLIGO sensitivity curve of Fig.~\ref{fig:minSearchDepth} over the given frequency range, while subsequent lines
illustrate slight variations when changing only one of the constraints (computing cost $\Cost_0$,
observation time $\Tobs$ or detectors, respectively).
In the last line of each table we also show a setup beating the $\fbar=2\nu$ torque-balance level
depth over the given frequency range assuming a 6-month science run using the ``Mid'' aLIGO
configuration, which is currently estimated~\cite{2013arXiv1304.0670L} to take place in about 2016-2017.
\begin{table*}[htbp]
\begin{ruledtabular}
  \caption{\label{tab:OptimalSetupsEaH-Demod}
    Optimal search setups and corresponding sensitivity estimates for a Scorpius~X-1 StackSlide
    search (with $\Tseg\le10\,\Days$)
    using the SFT-based demodulation algorithm, assuming negligible uncertainty
    $\Delta\ecc=0$ on the eccentricity.
    The SFT length assumed is $\Tsft(430~\Hz)=240\,\Sec$ and $\Tsft(250~\Hz)=320\,\Sec$
    (see Appendix~\ref{Sec:Tsft-choice}, for $\mis_\SFT=0.01$).
    The first line shows a setup beating the ``Final'' $\fbar=2\nu$ torque-balance
    depth (see Fig.~\ref{fig:minSearchDepth}) over the given frequency range, and subsequent lines
    illustrate the results for slightly different constraints on computing cost $\Cost_0$,
    observation time $\Tobs$ and detectors (IFOs).
    The last line shows a setup beating the $\fbar=2\nu$ torque-balance level depth over the
    given frequency range with a ``Mid'' aLIGO configuration.
    In all cases the template-bank dimensions are found as $\coh{n}=3$ and $\inc{n}=4$.
    The computing-cost unit ``$\mathrm{EM}$'' corresponds to running on Einstein@Home for a month, and
    is explained in more detail in Sec.~\ref{sec:computing-cost-model}.
  }
\begin{tabular}{c c c c || c c c c c c c || c}
$\fbar$ & $\Cost_0$ &  IFOs & $\Tobs$   & $\Nseg$ & $\Tseg$   & $\misMaxCoh$ & $\misMaxInc$ & $\NtCoh$ & $\NtInc$ & $\CCoh/\CInc$ & $\SensDepth{90\%}{\mathrm{1e-10}}$ \\
(Hz)     & (EM)       &       & $(\Days)$ &          & $(\Days)$ &               &               &           &           &                 & $(1/\sqrt{\Hz})$ \\
\hline
\hline
$[20,430]$ & $12$ &  HL & 360.0 &  $43$ & $8.30$ &$0.71$ &$0.04$ &$3.4{\times}10^{13}$ & $7.5{\times}10^{16}$ & $22$ & $100$\\
\hline
$[20,430]$ & $6$ &  HL & 360.0 &  $83$ & $4.36$ &$0.74$ &$0.06$ &$1.7{\times}10^{13}$ & $2.4{\times}10^{16}$ & $18$ & $86$\\
$[20,430]$ & $12$ &  HL & 180.0 &  $18$ & $10.00$ &$0.50$ &$0.02$ &$6.9{\times}10^{13}$ & $1.5{\times}10^{17}$ & $27$ & $91$\\
$[20,430]$ & $12$ &  HLV & 360.0 &  $62$ & $5.77$ &$0.73$ &$0.04$ &$2.3{\times}10^{13}$ & $5.1{\times}10^{16}$ & $22$ & $112$\\
\hline
\multicolumn{12}{l}{6 months science run with ``Mid'' aLIGO:}\\
\hline
\hline
$[40,230]$ & $6$ &  HL & 180.0 &  $18$ & $10.00$ &$0.18$ &$0.010$ &$4.8{\times}10^{13}$ & $7.8{\times}10^{16}$ & $26$ & $102$\\
\end{tabular}

  \caption{\label{tab:OptimalSetupsEaH-ResampCirc}
    Optimal search setups and corresponding sensitivity estimates for a Scorpius~X-1 StackSlide search
    (with $\Tseg\le10\,\Days$) using the FFT-based resampling algorithm, assuming negligible uncertainty $\Delta\ecc=0$ on the
    eccentricity.
    In all cases the template-bank dimensionalities are found as $\coh{n}=3$ and $\inc{n}=4$.
    The notation is the same as in Table~\ref{tab:OptimalSetupsEaH-Demod}.
  }
\begin{tabular}{c c c c || c c c c c c c || c}
$\fbar$ & $\Cost_0$ &  IFOs & $\Tobs$   & $\Nseg$ & $\Tseg$   & $\misMaxCoh$ & $\misMaxInc$ & $\NtCoh$ & $\NtInc$ & $\CCoh/\CInc$ & $\SensDepth{90\%}{\mathrm{1e-10}}$ \\
(Hz)     & (EM)       &       & $(\Days)$ &          & $(\Days)$ &               &               &           &           &                 & $(1/\sqrt{\Hz})$ \\
\hline
\hline
$[20,630]$ & $12$ &  HL & 360.0 &  $36$ & $10.00$ &$0.04$ &$0.03$ &$1.1{\times}10^{16}$ & $8{\times}10^{17}$ & $1.6$ & $129$\\
\hline
$[20,630]$ & $6$ &  HL & 360.0 &  $36$ & $10.00$ &$0.06$ &$0.05$ &$5.5{\times}10^{15}$ & $3.8{\times}10^{17}$ & $1.7$ & $127$\\
$[20,630]$ & $1$ &  HL & 360.0 &  $36$ & $10.00$ &$0.19$ &$0.12$ &$9.7{\times}10^{14}$ & $5.6{\times}10^{16}$ & $2.1$ & $117$\\
$[20,630]$ & $12$ &  HL & 180.0 &  $18$ & $10.00$ &$0.03$ &$0.02$ &$2.5{\times}10^{16}$ & $1.4{\times}10^{18}$ & $2.1$ & $105$\\
$[20,630]$ & $12$ &  HLV & 360.0 &  $36$ & $10.00$ &$0.05$ &$0.03$ &$7.5{\times}10^{15}$ & $7.2{\times}10^{17}$ & $1.9$ & $155$\\
\hline
\multicolumn{12}{l}{6 months science run with ``Mid'' aLIGO:}\\
\hline
\hline
$[40,250]$ & $6$ &  HL & 180.0 &  $18$ & $10.00$ &$0.006$ &$0.004$ &$1.1{\times}10^{16}$ & $6.9{\times}10^{17}$ & $2$ & $107$\\
\end{tabular}

  \caption{\label{tab:OptimalSetupsEaH-ResampEcc}
    Optimal search setups and corresponding sensitivity estimates for a Scorpius~X-1 StackSlide search
    (with $\Tseg\le10\,\Days$) using the FFT-based resampling method, assuming a pessimistic uncertainty $\Delta\ecc=0.087$ on
    eccentricity (see Table~\ref{tab:Scorpius-X1MDCvalues}).
    In all cases the template-bank dimensionalities are found as $\coh{n}=5$ and $\inc{n}=6$.
    The notation is the same as in Table~\ref{tab:OptimalSetupsEaH-Demod}.
    }
 \begin{tabular}{c c c c || c c c c c c c || c}
$\fbar$ & $\Cost_0$ &  IFOs & $\Tobs$   & $\Nseg$ & $\Tseg$   & $\misMaxCoh$ & $\misMaxInc$ & $\NtCoh$ & $\NtInc$ & $\CCoh/\CInc$ & $\SensDepth{90\%}{\mathrm{1e-10}}$ \\
(Hz)     & (EM)       &       & $(\Days)$ &          & $(\Days)$ &               &               &           &           &                 & $(1/\sqrt{\Hz})$ \\
\hline
\hline
$[20,200]$ & $12$ &  HL & 360.0 &  $36$ & $10.00$ &$1.18$ &$0.54$ &$1.3{\times}10^{16}$ & $5.7{\times}10^{17}$ & $2.6$ & $60$\\
\hline
$[20,200]$ & $6$ &  HL & 360.0 &  $36$ & $10.00$ &$1.56$ &$0.68$ &$6.3{\times}10^{15}$ & $2.8{\times}10^{17}$ & $2.7$ & $55$\\
$[20,200]$ & $12$ &  HL & 180.0 &  $18$ & $10.00$ &$0.88$ &$0.35$ &$2.6{\times}10^{16}$ & $10^{18}$ & $3$ & $56$\\
$[20,200]$ & $12$ &  HLV & 360.0 &  $36$ & $10.00$ &$1.37$ &$0.55$ &$8.6{\times}10^{15}$ & $5.2{\times}10^{17}$ & $3$ & $68$\\
\hline
\multicolumn{12}{l}{6 months science run with ``Mid'' aLIGO:}\\
\hline
\hline
$[60,140]$ & $6$ &  HL & 180.0 &  $18$ & $10.00$ &$0.57$ &$0.21$ &$1.3{\times}10^{16}$ & $5{\times}10^{17}$ & $3.1$ & $74$\\
\end{tabular}
\end{ruledtabular}
\end{table*}

\begin{table*}[htbp]
\begin{ruledtabular}
  \caption{\label{tab:OptimalSetupsEaH-Demod_v2}
    Optimal search setups and corresponding sensitivity estimates for a Scorpius~X-1 StackSlide search
    (with $\Tseg\le3\,\Days$) using the SFT-based demodulation algorithm, assuming negligible uncertainty
    $\Delta\ecc=0$ on the eccentricity.
    The SFT length assumed is $\Tsft(420\Hz)=250\,\Sec$ and $\Tsft(150\Hz)=410\,\Sec$
    (see Appendix~\ref{Sec:Tsft-choice}, for $\mis_\SFT=0.01$).
    The first line shows a setup beating the ``Final'' $\fbar=2\nu$ torque-balance
    depth (see Fig.~\ref{fig:minSearchDepth}) over the given frequency range, and subsequent lines
    illustrate the results for slightly different constraints on computing cost $\Cost_0$,
    observation time $\Tobs$ and detectors (IFOs).
    The last line shows a setup beating the $\fbar=2\nu$ torque-balance level depth over the
    given frequency range with a ``Mid'' aLIGO configuration.
    In all cases the template-bank dimensions are found as $\coh{n}=3$ and $\inc{n}=4$.
  }
 \begin{tabular}{c c c c || c c c c c c c || c}
$\fbar$ & $\Cost_0$ &  IFOs & $\Tobs$   & $\Nseg$ & $\Tseg$   & $\misMaxCoh$ & $\misMaxInc$ & $\NtCoh$ & $\NtInc$ & $\CCoh/\CInc$ & $\SensDepth{90\%}{\mathrm{1e-10}}$ \\
(Hz)     & (EM)       &       & $(\Days)$ &          & $(\Days)$ &               &               &           &           &                 & $(1/\sqrt{\Hz})$ \\
\hline
\hline
$[20,420]$ & $12$ &  HL & 360.0 &  $120$ & $3.00$ &$0.34$ &$0.03$ &$3.5{\times}10^{13}$ & $4.3{\times}10^{16}$ & $14$ & $94$\\
\hline
$[20,420]$ & $6$ &  HL & 360.0 &  $120$ & $3.00$ &$0.54$ &$0.05$ &$1.8{\times}10^{13}$ & $2{\times}10^{16}$ & $15$ & $87$\\
$[20,420]$ & $12$ &  HL & 180.0 &  $60$ & $3.00$ &$0.21$ &$0.02$ &$7.1{\times}10^{13}$ & $7.4{\times}10^{16}$ & $16$ & $80$\\
$[20,420]$ & $12$ &  HLV & 360.0 &  $120$ & $3.00$ &$0.44$ &$0.04$ &$2.4{\times}10^{13}$ & $3.6{\times}10^{16}$ & $16$ & $110$\\
\hline
\multicolumn{12}{l}{6 months science run with ``Mid'' aLIGO:}\\
\hline
\hline
$[50,150]$ & $6$ &  HL & 180.0 &  $60$ & $3.00$ &$0.03$ &$0.003$ &$5.7{\times}10^{13}$ & $4.3{\times}10^{16}$ & $14$ & $85$\\
\end{tabular}

  \caption{\label{tab:OptimalSetupsEaH-ResampCirc_v2}
    Optimal search setups and corresponding sensitivity estimates for a Scorpius~X-1 StackSlide search
    (with $\Tseg\le3\,\Days$) using the FFT-based resampling algorithm, assuming negligible uncertainty $\Delta\ecc=0$ on the
    eccentricity.
    In all cases the template-bank dimensionalities are found as $\coh{n}=3$ and $\inc{n}=4$.
    The notation is the same as in Table~\ref{tab:OptimalSetupsEaH-Demod}.
  }
 \begin{tabular}{c c c c || c c c c c c c || c}
$\fbar$ & $\Cost_0$ &  IFOs & $\Tobs$   & $\Nseg$ & $\Tseg$   & $\misMaxCoh$ & $\misMaxInc$ & $\NtCoh$ & $\NtInc$ & $\CCoh/\CInc$ & $\SensDepth{90\%}{\mathrm{1e-10}}$ \\
(Hz)     & (EM)       &       & $(\Days)$ &          & $(\Days)$ &               &               &           &           &                 & $(1/\sqrt{\Hz})$ \\
\hline
\hline
$[20,500]$ & $12$ &  HL & 360.0 &  $120$ & $3.00$ &$0.02$ &$0.02$ &$3.2{\times}10^{15}$ & $2.4{\times}10^{17}$ & $1.6$ & $102$\\
\hline
$[20,500]$ & $6$ &  HL & 360.0 &  $120$ & $3.00$ &$0.04$ &$0.03$ &$1.6{\times}10^{15}$ & $1.1{\times}10^{17}$ & $1.7$ & $101$\\
$[20,500]$ & $1$ &  HL & 360.0 &  $120$ & $3.00$ &$0.12$ &$0.07$ &$2.9{\times}10^{14}$ & $1.7{\times}10^{16}$ & $2.1$ & $96$\\
$[20,500]$ & $12$ &  HL & 180.0 &  $60$ & $3.00$ &$0.01$ &$0.01$ &$6.7{\times}10^{15}$ & $4.4{\times}10^{17}$ & $1.8$ & $84$\\
$[20,500]$ & $12$ &  HLV & 360.0 &  $120$ & $3.00$ &$0.03$ &$0.02$ &$2.3{\times}10^{15}$ & $2.1{\times}10^{17}$ & $1.9$ & $123$\\
\hline
\multicolumn{12}{l}{6 months science run with ``Mid'' aLIGO:}\\
\hline
\hline
$[50,190]$ & $6$ &  HL & 180.0 &  $60$ & $3.00$ &$0.003$ &$0.002$ &$3.4{\times}10^{15}$ & $2.1{\times}10^{17}$ & $2$ & $85$\\
\end{tabular}

  \caption{\label{tab:OptimalSetupsEaH-ResampEcc_v2}
    Optimal search setups and corresponding sensitivity estimates for a Scorpius~X-1 StackSlide search
    (with $\Tseg\le3\,\Days$) using the FFT-based resampling method, assuming a pessimistic uncertainty $\Delta\ecc=0.087$ on
    eccentricity (see Table~\ref{tab:Scorpius-X1MDCvalues}).
    In all cases the template-bank dimensionalities are found as $\coh{n}=5$ and $\inc{n}=6$.
    The notation is the same as in Table~\ref{tab:OptimalSetupsEaH-Demod}.
    }
  \begin{tabular}{c c c c || c c c c c c c || c}
$\fbar$ & $\Cost_0$ &  IFOs & $\Tobs$   & $\Nseg$ & $\Tseg$   & $\misMaxCoh$ & $\misMaxInc$ & $\NtCoh$ & $\NtInc$ & $\CCoh/\CInc$ & $\SensDepth{90\%}{\mathrm{1e-10}}$ \\
(Hz)     & (EM)       &       & $(\Days)$ &          & $(\Days)$ &               &               &           &           &                 & $(1/\sqrt{\Hz})$ \\
\hline
\hline
$[20,160]$ & $12$ &  HL & 360.0 &  $120$ & $3.00$ &$0.76$ &$0.34$ &$3.8{\times}10^{15}$ & $1.7{\times}10^{17}$ & $2.6$ & $58$\\
\hline
$[20,160]$ & $6$ &  HL & 360.0 &  $120$ & $3.00$ &$1.00$ &$0.44$ &$1.9{\times}10^{15}$ & $8.3{\times}10^{16}$ & $2.7$ & $50$\\
$[20,160]$ & $12$ &  HL & 180.0 &  $60$ & $3.00$ &$0.57$ &$0.22$ &$7.8{\times}10^{15}$ & $3.1{\times}10^{17}$ & $3$ & $59$\\
$[20,160]$ & $12$ &  HLV & 360.0 &  $120$ & $3.00$ &$0.88$ &$0.35$ &$2.6{\times}10^{15}$ & $1.6{\times}10^{17}$ & $3$ & $65$\\
\hline
\multicolumn{12}{l}{6 months science run with ``Mid'' aLIGO:}\\
\hline
\hline
$[70,110]$ & $6$ &  HL & 180.0 &  $60$ & $3.00$ &$0.34$ &$0.13$ &$3.9{\times}10^{15}$ & $1.5{\times}10^{17}$ & $3.1$ & $71$\\
\end{tabular}
\end{ruledtabular}
\end{table*}

Summarizing these results, we find that the SFT-based demodulation method can only reach the torque-balance level
assuming we can neglect the uncertainty in eccentricity ($\Delta\ecc=0$), shown in
Tables~\ref{tab:OptimalSetupsEaH-Demod} and \ref{tab:OptimalSetupsEaH-Demod_v2}.
The FFT-based resampling method, in the case of $\Delta\ecc=0$, can beat the torque-balance level over a wider frequency
range up to $\sim630\,\Hz$ (for $\Tseg\le10\,\Days$) or up to $\sim500\,\Hz$ (for
$\Tseg\le3\,\Days$), even at substantially reduced computing cost
(cf.\ Tables~\ref{tab:OptimalSetupsEaH-ResampCirc} and \ref{tab:OptimalSetupsEaH-ResampCirc_v2}),
and can still beat the torque-balance limit (albeit only up to a smaller frequency of
$\sim160\,\Hz-200\,\Hz$, depending on the constraint on $\Tseg$) in the case of
substantial uncertainty ($\Delta\ecc=0.087$) on eccentricity, as seen in
Tables~\ref{tab:OptimalSetupsEaH-ResampEcc} and \ref{tab:OptimalSetupsEaH-ResampEcc_v2}.

At a lower-false alarm probability of $\pFA=10^{-14}$, corresponding to a level at which
one might set upper limits without any further followup, the corresponding sensitivity depth
$\SensDepth{90\%}{1e-14}$ would be roughly $10\%$ lower compared to the values of
$\SensDepth{90\%}{1e-10}$ quoted in
Tables~\ref{tab:OptimalSetupsEaH-Demod}--~\ref{tab:OptimalSetupsEaH-ResampEcc_v2}.

Note that some of the optimal setups include maximal mismatches $\misMax$ of order unity, which
would violate the metric approximation discussed in Sec.~\ref{sec:templ-banks-metr}.
In order to fully quantify this effect, further Monte-Carlo tests on the higher-order metric
deviations would be required, but qualitatively we know from previous results
[e.g.\ see \cite{Prix:2006wm,Wette:2013wza}, Figs.~\ref{fig:testing_CO_LS}~(f),~\ref{fig:testing_SC_LS}~(f),
and metric tests we performed at
$\misMax=0.5$ that are not shown here] that in this regime the metric approximation
generally tends to \emph{overestimate} the actual measured mismatches.
Furthermore, the distribution of sampled mismatches will be peaked around their mean value, which
for $\Ans{4}$ lattices is $\avgT{\mis} \approx 0.5\,\misMax$, and will therefore not be affected as much
by these deviations.

\subsection{Comparison to previous sensitivities}
\label{sec:comp-prev-sens}

In order to put these sensitivities in context, we consider some previously-achieved
sensitivities for Scorpius~X-1 CW searches.
The first coherent Scorpius~X-1 search presented in~\cite{Abbott:2006vg}
was computationally limited to using a total of $\sim9\,\Hours$ of data collected
by LIGO during the second science run with the Livingston and Hanford detectors.
Such a search achieved upper limits corresponding to a sensitivity depth of
approximately $\SensDepth{95\%}{}\approx 4\,\Hz^{-1/2}$, which is roughly consistent with the
theoretical expectation~\cite{Wette:2011eu} of $\SensDepth{95\%}{1e-14}\approx 4.5\,\Hz^{-1/2}$ for a
fully-coherent two-detector search.
The recent search on the fifth LIGO science run, which uses a semi-coherent sideband method \cite{2014arXiv1412.0605T},
achieved a substantially improved sensitivity depth of $\SensDepth{95\%}{}\approx 30\,\Hz^{-1/2}$
over a frequency band of $50-550\,\Hz$.
Note that the software-injected signals used in the recent Scorpius~X-1 MDC \cite{ScoX1:MDC1} are
found (over a frequency range of $50-1500\,\Hz$) in the depth range
$\SensDepth{}{} \in [2.6,\, 104]\,\Hz^{-1/2}$ for the ``open'' injections, and
$\SensDepth{}{} \in [2.4,\, 58]\,\Hz^{-1/2}$ for the ``closed'' injections.

\section{Conclusions}
\label{sec:conc}

The work presented here consists of three main parts.

In the first part we re-derived the (coherent and
semi-coherent) metric expressions for the long-segment and short-segment regimes, assuming
a low-eccentricity binary orbit. We found that (as has also been noted earlier), the long-segment
regime does require refinement in the orbital angular velocity $\Om$, contrary to an earlier
result in the literature \cite{Messenger:2011rg}.
We have extended these findings to allow for general offsets between the orbital reference time epoch
$\tAsc$ and the observation epoch, which explicitly show that there is generally \emph{no} increase in
computing cost with increasing time passed since the original orbital reference epoch if both $\Om$
and $\tAsc$ are fully resolved by the metric.

In the second part we subjected the analytic metric expressions to extensive Monte-Carlo tests
comparing their predictions against measured SNR loss in software-injection studies, and found robust
agreement within their range of applicability.

In the final part of this work we used the metric template counts to semi-analytically estimate the
optimal achievable StackSlide sensitivity of a directed search for Scorpius~X-1. We found that the
predicted torque-balance limit could be reachable for the first time with an Einstein@Home search
using data from a $6$-month ``Mid'' aLIGO run (currently planned for 2016-2017), and should be beatable
over a substantial frequency range in ``Final'' aLIGO. However, the frequency range over which the
torque-balance limit can be beaten depends strongly on the Scorpius~X-1 parameter-space uncertainty, most
notably the orbital eccentricity and the assumed characteristics of spin wandering.

More work is required to better understand the modelling and effects of spin wandering, as well as the
resulting constraints on search methods. It would seem desirable to develop a more robust
statistic that properly takes spin wandering into account, for example by marginalizing
over the spindown uncertainty in each segment before incoherently summing them.

\section*{Acknowledgments}
We would like to thank Karl Wette, Chris Messenger, John T.~Whelan, David Keitel, Badri Krishnan,
Maria Alessandra Papa and Holger Pletsch for useful discussions.
We further thank Sammanani Premachandra and Duncan Galloway for providing us with their as-yet
unpublished parameter estimate on the eccentricity of Scorpius~X-1.
PL thanks the support of the \textit{Sonderforschungsbereich Transregio} (SFB/TR7)
collaboration and the University of Rome ``Sapienza''.
Numerical simulations were performed on the ATLAS computer cluster of the
\textit{Max-Planck-Institut f\"ur Gravitationsphysik}.
This work was completed using exclusively FreeSoftware tools: the algebra was checked using
Maxima, the numerical scripting was done in Octave with SWIG bindings (provided by Karl
Wette) to LALSuite, and all plots were prepared using Octave/Gnuplot.
This paper has been assigned document number LIGO-P1500006-v3.

\appendix

\section{Inverting the $\uCoord$-coordinates back into physical coordinates}
\label{sec:invert-u-coord}

Let us define re-scaled coordinates
\begin{equation}
  \label{eq:49}
  w_k \equiv \frac{u_k}{\Om^k}\,,
\end{equation}
and consider the set of Eqs.~\eqref{eq:ucoords} for
$w_k = w_k(\Dop)$.

\subsection{General 6D elliptic case}
\label{sec:general-6d-elliptic}

Let us take linear combinations of the equations for $w_2,\ldots w_5$ to express
\begin{align}
  \sin\psim &= \frac{1}{3\fbar\asini}\,(w_4 + 4 w_2)\,,  \label{eq:50a}\\
  \cos\psim &= \frac{1}{3\fbar\asini}\,(w_5 + 4 w_3)\,, \label{eq:50b}\\
  \sin2\psim&= -\frac{1}{12\fbar\asini\,\ecc^2} [2\kappa (w_2 + w_4) + \eta (w_3 + w_5)]\,,\label{eq:50c}\\
  \cos\psim&= -\frac{1}{12\fbar\asini\,\ecc^2} [\kappa (w_3 + w_5) - 2\eta (w_2 + w_4)]\,,\label{eq:50d}
\end{align}
where we used $\ecc^2 = \kappa^2 + \eta^2$. We now insert these into the equation for $w_6$ to obtain
\begin{equation}
  \label{eq:51}
  w_6 + 5 w_4 + 4 w_2 = 0\,,
\end{equation}
which yields a quadratic equation for $\Om^2$, with solution
\begin{equation}
  \label{eq:52}
  \Om^2 = -\frac{5 u_4}{8 u_2} + \sqrt{ \left(\frac{5 u_4}{8 u_2}\right)^2 - \frac{u_6}{4 u_2}}\,,
\end{equation}
which requires $25 u_4^2 - 16 u_2 u_6 > 0$ for real-valued $\Om^2$.
Note that once we know $\Om$, then all $w_k$ are known numerically as well.
Inserting Eqs.~\eqref{eq:50a}-\eqref{eq:50d} into the equation for $w_1$ brings us to find
\begin{equation}
  \label{eq:53}
  \freq = u_1 + \frac{5 u_3}{\Om^2} + \frac{u_5}{4\Om^4}\,.
\end{equation}
Using $\sin^2\psim+\cos^2\psim=1$, we obtain
\begin{equation}
  \label{eq:54}
  3\fbar\asini = \sqrt{ (w_5 + 4 w_3)^2 + (w_4 + 4 w_2)^2 }\,,
\end{equation}
where we note that for constant-frequency signals, we numerically have $\fbar=\freq$.
Hence, given $\freq$ from the previous equation, we easily get $\asini$.
Proceeding similarly for $\sin^22\psim + \cos^22\psim=1$, we find
\begin{equation}
  \label{eq:55}
  \ecc = \frac{1}{12\fbar\asini}\sqrt{ (w_3 + w_5)^2 + 4(w_2 + w_4)^2}\,.
\end{equation}
We can simply invert Eqs.~\eqref{eq:50a},\eqref{eq:50b} to obtain $\psim$ and therefore $\tAsc$.
Finally, solving the linear system of equations ${w_2,w_3}$ for $\kappa,\eta$, yields
\begin{align}
  \label{eq:56}
  \kappa &= \frac{1}{4\fbar\asini}\,\left[ 2(w_2 - \fbar\asini \sin\psim) \sin2\psim  \right.\nonumber\\
  & \qquad \left. {} + (w_3 - \fbar\asini\cos\psim) \cos2\psim \right]\,,\\
  \eta &= \frac{1}{4\fbar\asini}\,\left[ (w_3 - \fbar\asini \cos\psim) \sin2\psim  \right.\nonumber\\
  & \qquad \left. {} + 2(\fbar\asini\sin\psim - w_2) \cos2\psim \right]\,,
\end{align}
which, using Eq.~\eqref{eq:55} and Eq.~\eqref{eq:26} gives us $\argp$.

\subsection{Special 4D circular case}
\label{sec:special-4d-circular}

In the special circular case, we can obtain a simpler solution by inverting the equations for
$\{w_1,\ldots w_4\}$ and setting $\kappa=\eta=0$, namely
\begin{align}
  \label{eq:57}
  w_1 &= \frac{f}{\Om} - \fbar\asini\cos\psim\,,\\
  w_2 &= \fbar\asini\,\sin\psim\,,\\
  w_3 &= \fbar\asini\,\cos\psim\,,\\
  w_4 &= -\fbar\asini\,\sin\psim\,.
\end{align}
By summing equations for $w_2$ and $w_4$, we find $w_2 + w_4 = 0$, which is a quadratic equation for
$\Om$ with solution
\begin{equation}
  \label{eq:58}
  \Om = \sqrt{-\frac{u_4}{u_2}}\,,
\end{equation}
which numerically determines all $w_k$. We can further see that
\begin{equation}
  \label{eq:59}
  \fbar\asini = \sqrt{w_2^2+w_3^2}\,,
\end{equation}
and
\begin{equation}
  \label{eq:60}
  \tan\psim = \frac{w_2}{w_3}\,,
\end{equation}
yielding $\tAsc$, and finally
\begin{equation}
  \label{eq:61}
  f = \fbar\asini\Om \cos\psim + u_1\,.
\end{equation}

\section{Maximal Doppler shift due to orbital motion}
\label{sec:maxim-doppl-shift}

Sometimes it is important to estimate the maximal Doppler shift the intrinsic signal frequency of a
binary CW signal can undergo due to orbital motion. From the phase model of Eq.~\eqref{eq:23} we
see that the instantaneous Doppler shift is
\begin{align}
  \label{eq:93}
  \left|\frac{d\Phase/d t}{2\pi\freq}-1\right| &= \left|\frac{d R}{c\,d t}\right| \nonumber\\
  &= \asini\Om \left|\frac{ \sqrt{1-\ecc^2}\cos{E}\cos\argp - \sin{E} \sin\argp}{1 - \ecc\cos{E}}\right|\nonumber\\
  & \le \asini\Om \frac{|\sin{E}\sin\argp| + |\cos{E}\cos\argp|}{|1 -\ecc\cos{E}|}\,,
\end{align}
where we used the fact that $|a+b|\le|a|+|b|$ and $\sqrt{1-\ecc^2}\le1$.
We further observe that
\begin{align}
  \label{eq:94}
  &|\cos{E}\cos\argp| + |\sin{E}\sin\argp| = \nonumber\\
  &\hspace*{0.3cm}\max\{|\cos(E+\argp)|,\, |\cos(E-\argp)|\} \le 1\,,
\end{align}
and $|1 - \ecc\cos{E}|\ge 1-\ecc$ to obtain
\begin{equation}
  \label{eq:95}
  \left|\frac{d\Phase/d t}{2\pi\freq}-1\right| \le \frac{\asini\Om}{1 - \ecc}\,.
\end{equation}

\section{Maximal SFT length}
\label{Sec:Tsft-choice}

By using the SFT-based demodulation method to compute the $\Fcoh$-statistic, the
computing cost per template increases linearly with the number of SFTs used [see
Eq.~\eqref{eq:Ccoarse}]. On the other hand, the maximal length of the SFT is limited by the
linear-phase approximation over the duration of each SFT.
In order to minimize the computing cost of this method, we therefore want to choose the longest
possible SFT duration $\Tsft$ [see Eq.~\eqref{eq:72}] with an acceptable error in the linear-phase approximation.

\begin{figure}[htbp]
  \centering
  \includegraphics[width=\columnwidth]{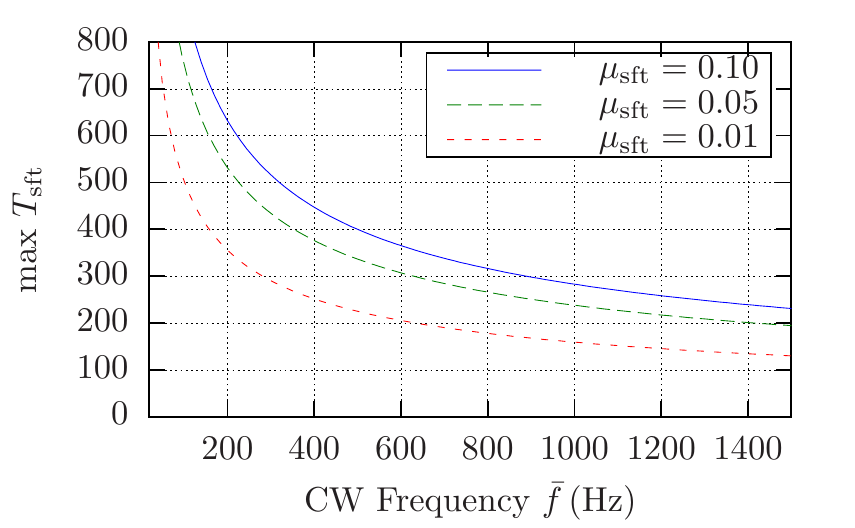}
  \caption{Maximal allowed SFT length $\Tsft$ of Eq.~\eqref{eq:70} for a Scorpius~X-1 demodulation
    $\Fcoh$-statistic search as function of frequency $\fbar$, for three different tolerated
    maximal mismatches $\mis_\SFT$.}
  \label{fig:maxTsft}
\end{figure}

In order to estimate the maximal phase-error of the linear-phase approximation over an SFTs, we can
conveniently re-use the short-segment limit expressions, and simply estimate the phase error as
\begin{equation}
  \label{eq:65}
  |\Delta\Phase| \sim |v_2| = 2\pi\, \frac{|u_2|}{2!}\left(\frac{\Tsft}{2}\right)^2
  \sim \frac{\pi}{4}\,\asini\,\fbar\,\Om^2\, \Tsft^{2}\,,
\end{equation}
and the corresponding mismatch is given via Eq.~\eqref{eq:46} as $\mis_\SFT \sim \coh{g}^{\ShortSeg,{v}}_{2 2}  v_2^2 =\frac{4}{45} v_2^2$.
Turning this around to express the maximal $\Tsft$ for a given maximal mismatch $\mis_\SFT$, we
obtain
\begin{equation}
  \label{eq:70}
  \Tsft^2(\fbar) \le \frac{ 6\sqrt{5\,\mis_\SFT} }{\pi\,\asini\,\fbar\,\Om^2}\,,
\end{equation}
where the \emph{largest} values of the parameter space being searched should be used for $\asini,\Om$ and $\fbar$.
This constraint on $\Tsft$ is illustrated in Fig.~\ref{fig:maxTsft} as a function of search
frequency $\fbar$ for Scorpius~X-1 parameters of Table~\ref{tab:Scorpius-X1MDCvalues}.
We see that for Scorpius~X-1 this limit is more stringent than the analogous constraint coming
purely from the Doppler effect due to the detector motion, which typically results in a choice of
$\Tsft=1800\,$s in searches for isolated CW sources.

\section{Optimal StackSlide solution for degenerate computing cost function}
\label{sec:optim-stacksl-solut}

In the case of the FFT-based resampling algorithm for the $\Fcoh$-statistic of Eq.~\eqref{eq:74}, we
encounter a degeneracy in the computing-cost function that had not been considered in the original
optimization study of \cite{Prix:2012yu}. Namely, when the search dimension $\Om$ is resolved in the
semi-coherent template bank, but unresolved in the per-template coherent banks, then we see from
Eqs.~\eqref{eq:31N}--\eqref{eq:86} that the number of templates (both for including
eccentricity $\coh{n}=5,\inc{n}=6$, and for the circular case $\coh{n}=3,\inc{n}=4$), scale as
$\coh{\Nt}\propto\Tseg$ and $\inc{\Nt}\propto\Nseg\Tseg$, respectively, where we assumed the ideal gapless case with $\refine=\Nseg$.
From the computing-cost expressions Eq.~\eqref{eq:Ccoarse} and Eq.~\eqref{eq:Cfine} we see that
therefore the computing-cost functions take the form
$\coh{\Cost}\propto \Nseg\Tseg \propto \Tobs$ and $\inc{\Cost}\propto(\Nseg\Tseg)^2\propto\Tobs^2$.
In this case the power-law coefficients in the formalism of \cite{Prix:2012yu} are
$\coh{\eta}=\coh{\delta}=1$ and $\inc{\eta}=\inc{\delta}=2$ (the resulting $\Nseg$ coefficients at
fixed $\Tobs$ are therefore $\coh{\varepsilon}=\inc{\varepsilon}=0$), which corresponds to a
degenerate case that has not been analyzed previously.
In this case one can only find a solution by constraining $\Tobs$. However, maximization of
sensitivity at fixed computing cost would then result in $\Nseg\rightarrow1$, i.e.\ a
fully coherent search (except if by increasing $\Tseg$ the template bank eventually starts to
resolve $\Om$ and therefore breaks the degeneracy).
In cases such as the Scorpius~X-1 search considered here, there is an astrophysically-motivated upper
bound on the segment length $\Tseg\le 10\,\Days$, and we therefore need to express the optimal
solution for constraints on both $\Tobs$ and $\Tseg$.
With $\Nseg$ and $\Tobs$ fixed, we can only optimize sensitivity over the respective template-bank
mismatches $\misMaxCoh$ and $\misMaxInc$.
This is achieved simply by solving Eq.~(91) of \cite{Prix:2012yu}, i.e.
$(\misMaxCoh/\coh{n})/(\misMaxInc/\inc{n})=\coh{\Cost}/\inc{\Cost}$, together with
$\coh{\Cost}+\inc{\Cost}=\Cost_0$ for $\misMaxInc$ and $\misMaxCoh$.

\bibliography{nr_MismatchForBinaryMetric}

\end{document}